\documentclass[sigconf, screen]{sty/acmart}

\usepackage{makecell}
\usepackage{graphicx}
\usepackage{xcolor}

\usepackage{colortbl}

\usepackage{graphicx}

\usepackage{listings}
\usepackage{multirow}
\usepackage{graphicx}

\lstdefinelanguage{Solidity}{
	keywords=[1]{anonymous, assembly, assert, balance, break, call, callcode, case, catch, class, constant, continue, constructor, contract, debugger, default, delegatecall, delete, do, else, emit, event, experimental, export, external, false, finally, for, function, gas, if, implements, import, in, indexed, instanceof, interface, internal, is, length, library, log0, log1, log2, log3, log4, memory, modifier, new, payable, pragma, private, protected, public, pure, push, require, return, returns, revert, selfdestruct, send, solidity, storage, struct, suicide, super, switch, then, this, throw, transfer, true, try, typeof, using, value, view, while, with, addmod, ecrecover, keccak256, mulmod, ripemd160, sha256, sha3}, 
	keywordstyle=[1]\color{blue}\bfseries,
	keywords=[2]{address, bool, byte, bytes, bytes1, bytes2, bytes3, bytes4, bytes5, bytes6, bytes7, bytes8, bytes9, bytes10, bytes11, bytes12, bytes13, bytes14, bytes15, bytes16, bytes17, bytes18, bytes19, bytes20, bytes21, bytes22, bytes23, bytes24, bytes25, bytes26, bytes27, bytes28, bytes29, bytes30, bytes31, bytes32, enum, int, int8, int16, int24, int32, int40, int48, int56, int64, int72, int80, int88, int96, int104, int112, int120, int128, int136, int144, int152, int160, int168, int176, int184, int192, int200, int208, int216, int224, int232, int240, int248, int256, mapping, string, uint, uint8, uint16, uint24, uint32, uint40, uint48, uint56, uint64, uint72, uint80, uint88, uint96, uint104, uint112, uint120, uint128, uint136, uint144, uint152, uint160, uint168, uint176, uint184, uint192, uint200, uint208, uint216, uint224, uint232, uint240, uint248, uint256, var, void, ether, finney, szabo, wei, days, hours, minutes, seconds, weeks, years},	
	keywordstyle=[2]\color{teal}\bfseries,
	keywords=[3]{block, blockhash, coinbase, difficulty, gaslimit, number, timestamp, msg, data, gas, sender, sig, value, now, tx, gasprice, origin},	
	keywordstyle=[3]\color{violet}\bfseries,
	identifierstyle=\color{black},
	sensitive=true,
	comment=[l]{//},
	morecomment=[s]{/*}{*/},
	commentstyle=\color{gray}\ttfamily,
	morestring=[b]',
	morestring=[b]",
  moredelim=[is][\color{red}]{@@}{@@},
  moredelim=[is][\color{green}]{!!}{!!},
  commentstyle=\color{purple}\ttfamily,
}

\usepackage{amsopn,amssymb}
\usepackage{microtype,xspace,graphicx,fancyvrb,multirow}
\usepackage[T1]{fontenc}

\usepackage{fp}
\usepackage{siunitx}

\usepackage{amsthm}

\newtheorem{definition}{Definition} 

\usepackage[linesnumbered,ruled,vlined]{algorithm2e}
\usepackage{pifont}
\newcommand{\circled}[1]{\ding{\numexpr171+#1\relax}}
\usepackage{bbding}

\newcommand{\KL}[1]{}
\newcommand{\WB}[1]{}
\newcommand{\TODO}[1]{}
\newcommand{\BW}[1]{}

\fvset{fontsize=\scriptsize,xleftmargin=8pt,numbers=left,numbersep=5pt}

\input{fmt}

\def\Snospace~{\S{}}

\newif\ifdraft\drafttrue
\newif\ifnotes\notestrue
\ifdraft\else\notesfalse\fi

\input{glyphtounicode}
\newcolumntype{R}[1]{>{\raggedleft\let\newline\\\arraybackslash\hspace{0pt}}p{#1}}

\newcommand{\squishlist}{
\begin{itemize}[noitemsep,nolistsep]
  \setlength{\itemsep}{-0pt}
}
\newcommand{\squishend}{
  \end{itemize}
}

\newcommand{\PP}[1]{
\noindent{\bf \IfEndWith{#1}{.}{#1}{#1.}}
}

\newcommand{\boxbeg}{
\vspace{2px}
\noindent\begin{tabular}{|l|}\hline
\begin{minipage}{3.2in}
\vspace{2px}
\noindent
}

\newcommand{\boxend}{
\vspace{2px}
\end{minipage}\\ \hline
\end{tabular}
\vspace{-10pt}
}

\copyrightyear{2026}
\acmYear{2026}
\setcopyright{cc}
\setcctype{by}
\acmConference[ICSE '26]{2026 IEEE/ACM 48th International Conference on Software Engineering}{April 12--18, 2026}{Rio de Janeiro, Brazil}
\acmBooktitle{2026 IEEE/ACM 48th International Conference on Software Engineering (ICSE '26), April 12--18, 2026, Rio de Janeiro, Brazil}
\acmPrice{}
\acmDOI{10.1145/3744916.3787799}
\acmISBN{979-8-4007-2025-3/2026/04}

\begin{document}
\title[GenDetect]{GenDetect: Generalizing Reactive Detection for Resilience Against Imitative DeFi Attack Cascade}



\ifdefined\DRAFT
 \pagestyle{plain}
 \lhead{Rev.~\therev}
 \rhead{\thedate}
 \cfoot{\thepage\ of \pageref{LastPage}}
\fi

\settopmatter{authorsperrow=4}

\newcommand{\UMNaffil}{%
  \affiliation{%
    \institution{University of Minnesota}
    \city{Minneapolis}
    \state{MN}
    \country{USA}
  }%
}

\newcommand{\XJTUaffil}{%
  \affiliation{%
    \institution{Xi'an Jiaotong University}
    \city{Xi'an}
    \state{Shaanxi}
    \country{China}
  }%
}

\newcommand{\JHUaffil}{%
  \affiliation{%
    \institution{Johns Hopkins University}
    \city{Baltimore}
    \state{Maryland}
    \country{USA}
  }%
}

\newcommand{\ZJUaffil}{%
  \affiliation{%
    \institution{Zhejiang University}
    \city{Hangzhou}
    \state{Zhejiang}
    \country{China}
  }%
}

\author{Bowen Cai}
\email{cai00254@umn.edu}
\orcid{0000-0002-2495-2108}
\UMNaffil

\author{Weiheng Bai}
\email{bai00093@umn.edu}
\UMNaffil

\author{Youshui Lu}
\email{yolu6176@uni.sydney.edu.au}
\XJTUaffil

\author{Haoran Xu}
\email{hxu65@jh.edu}
\JHUaffil

\author{Yuannan Yang}
\email{yyang181@jh.edu}
\JHUaffil

\author{Yajin Zhou}
\email{yajin_zhou@zju.edu.cn}
\ZJUaffil

\author{Kangjie Lu}
\email{kjlu@umn.edu}
\UMNaffil



\date{}
\begin{abstract}

As blockchain ecosystems grow, financially motivated attackers have increasingly exploited vulnerabilities in decentralized finance (DeFi) protocols, resulting in frequent and severe losses. Unlike conventional cyberattacks, DeFi exploits propagate rapidly due to the transparent and composable nature of smart contracts.  In this setting, we identify a critical behavioral pattern: \emph{Imitative Attack Cascade}, where an initial successful exploit is quickly followed by a flurry of mimicking transactions that reuse attack logic with minor modifications or parameter changes. Our empirical analysis shows that over 69\% of DeFi attacks exhibit strong behavioral similarity to earlier incidents, often occurring within hours or days of the initial attack.

This phenomenon highlights a fundamental limitation in current reactive detection workflows. While the initial attacks are often flagged through heuristic alerts, such as Tornado Cash traces, anomalous nonce usage, or known exploiter labels, these signals require manual validation and the construction of handcrafted detection rules through trace analysis. This process is labor-intensive and slow, resulting in unacceptable latency while follow-up attacks continue to spread.
Motivated by this gap, our research goal is to ensure that once an attack has been observed, even a single instance, it can be rapidly abstracted into an actionable and generalizable detection rule, enabling scalable protection against imitative attacks.

We decompose the problem into two core challenges: (I) abstracting the semantics of diverse, obscure function signatures, and (II) matching transaction logic in noisy, evasive traces. To address these, we leverage two key insights: (i) the open-source nature of most DeFi protocols enables high-fidelity semantic classification of function signatures; (ii) contract labels allow us to isolate essential logic by filtering irrelevant calls and classifying attack intent. Based on these, we develop a reactive detection framework, \textit{GenDetect}, which achieves strong benchmark performance (ACC: 98\%, FPR: 1\%, FNR: 3\%) and, critically, discovers 56 previously unrevealed attacks from the past three years, highlighting its practical effectiveness. Our source code and dataset are released on \href{https://github.com/NobodyIsAnonymous/GenDetect_ICSE2026}{Github}.

\end{abstract}

\maketitle
\sloppy

\section{Introduction}
The rapid growth of blockchain ecosystems has been accompanied by a sharp rise in security threats. Each year, hundreds of security incidents result in billions of dollars in losses~\cite{Cymetrics}. Among these, Decentralized Finance (DeFi)~\cite{moncada2021next, xu2022short, aquilina2024decentralized, gogol2024sok} stands out as the most prominent and lucrative target due to its high-value assets and composable smart contract design. The transparency and reusability of smart contracts make successful exploits particularly easy to replicate, contributing to the rapid proliferation of follow-up attacks.
Yet detection methods remain inadequate in the face of these escalating threats. Industry solutions~\cite{attack_survey} often rely on ad hoc heuristics--such as hardcoded exploiter addresses, Tornado Cash interactions~\cite{nadler2023tornado}, or timing-based indicators like nonce or contract creation time--that may help identify the first exploit but fail to capture its evolving variants. Academic approaches, on the other hand, typically focus on specific vulnerability types (e.g., price manipulation~\cite{wu2023defiranger, xi2024pomabuster, wang2021promutator, mo2023priceFeed}, flashloans~\cite{xia2023flashLoan}, or reentrancy~\cite{zhang2020txspector}), lacking the generality required to address real-world exploits at scale. As a result, once a vulnerability is exposed, it is often repeatedly exploited--sometimes for months--due to delays in designing and deploying effective detection patterns.




\PP{The Imitative DeFi Attack Cascade} A key motivation for our research lies in the empirical observations that most DeFi attacks are not isolated or fundamentally novel, but rather imitations, adaptations, or direct replications of previously disclosed exploits. In real-world incidents, once an attack is proven effective, it often triggers an \emph{imitative attack cascade}--sometimes reusing entire contracts with minimal modifications, and sometimes varying parameters or call sequences. To quantify this pattern, we analyze confirmed attack transactions from Phalcon~\cite{phalcon} and DeFiHackLab~\cite{defihacklab} across multiple protocols. Our analysis reveals that over 69\% of the attacks--collectively accounting for more than \$2 billion in losses--exhibit strong behavioral similarities to the earlier exploits. Some follow-ups occur within hours or days of the original, while others replicate nearly identical strategies over a year later. This persistence underscores not only the reusability of vulnerable patterns but also the absence of timely and generalizable defenses. Without a mechanism to convert each new exploit into a broadly applicable detection pattern, today's undetected attack becomes tomorrow's widely reused blueprint. Detection generalization is therefore not a supplementary feature--it is the core capability needed for sustainable defense.


\PP{Prior Work}
In practice, the industry relies on broad, heuristic-based signals--such as low-nonce initiators, Tornado Cash funding paths, or known exploiter addresses--to raise alerts~\cite{IndustryApproach}. Although these methods offer wide coverage and increase the odds of catching the first exploit, they suffer from extremely high false positives and require substantial manual effort to verify. More importantly, even when the first attack is confirmed, such approaches cannot automatically generate rules to detect follow-up \emph{imitative attacks}. This limitation stems from the fact that most existing detection systems follow a reactive loop of "discovery$\rightarrow$manual analysis$\rightarrow$rule generation", as seen in TxSpector~\cite{zhang2020txspector}, DeFiRanger~\cite{wu2023defiranger}, and POMABuster~\cite{xi2024pomabuster}. These tools manually extract behavioral signatures from traces to produce handcrafted rules, but their reliance on human intervention introduces unacceptable latency--especially in blockchain environments where imitation attacks can emerge within minutes. Alarmingly, certain vectors remain exploitable months or even years after disclosure~\cite{recur_1, recur_2, recur_idol_nft, recur_4}--not due to ignorance, but because current defense systems lack the capability to generalize and automate detection across related contexts.

This highlights the core value of our work: Rather than aiming to be the first to discover new attacks, we take the initial disclosure--identified through external alerts, public reports, or monitoring tools--as input. By capturing such cases through \textbf{generalizable semantic representations}, we can \textbf{rapidly and automatically produce detection strategies}, avoiding the delays and brittleness of handcrafted rules. While no system can anticipate entirely new categories in advance, our method ensures that once an incident is revealed, any imitative form of attack is unlikely to happen stealthily. In this way, we complement existing efforts by bridging the gap between discovery and timely, scalable detection for \emph{imitative attacks}.






\PP{Research Goal and Challenges} In this paper, our goal is to rapidly generalize an initial attack incident into a reactive detection to enable the precise resilience for \emph{imitative DeFi attacks}. 
We decompose the goal into two main challenges: (C1) Semantic Abstraction of Function Signatures. Transaction traces often contain diverse and complex function signatures. Existing word-level approaches, such as keyword matching or word-vector embeddings, struggle to capture their true semantics, especially in the context of DeFi-specific operations. (C2) Robust Matching of Transaction Logic. DeFi transaction traces are typically noisy, containing numerous benign operations or even intentionally obfuscated behaviors. This severely undermines the effectiveness of naive pattern-matching techniques like Longest Common Subsequence (LCS), which are sensitive to trace variations and cannot tolerate structural disguises. 

We leverage two key insights to address the challenges respectively:
(I) Source-based Semantic Extraction. We observe that most real-world attacks primarily target valuable DeFi protocols, which are typically open-sourced. This allows us to go beyond surface-level function names and directly utilize source code to classify function signatures based on their actual implementations. Compared to word-level recognition, our method yields more semantically accurate representations of function behavior.
(II) Contract label-based Logic Matching. We observe that a transaction's core intent often emerges from direct invocations by Externally Owned Accounts (EOAs) or affiliated contracts, particularly when these calls interact with two categories of targets: \circled{1} Core asset contracts (e.g., ETH, WBNB, USDT) for value extraction, and \circled{2} Protocol-specific tokens for exploitation. 
By isolating such initiators and grouping their interactions by target type, we reduce semantic noise and improve the robustness of logic matching against noisy even evasive behaviors.

\PP{Evaluation Summary}
We evaluate our system on both benchmark and real-world datasets, demonstrating strong performance across accuracy, generalizability, and real-time capability. On the DeFiHackLab~\cite{defihacklab} benchmark, our approach achieves a cross-validation F1-score of 98\%, with a false positive rate under 1\%--substantially lower than the 16\% reported by the state-of-the-art tool Forta~\cite{Forta}. Our ablation study further confirms the effectiveness of our design: Insight I contributes a 50\% relative improvement, and Insight II adds 9\%. In zero-shot evaluation on the Phalcon~\cite{phalcon} dataset, our method achieves 76\% accuracy, compared to 65\% by Forta. Beyond known benchmarks, our system identifies 56 (\autoref{sec:exp_RQ2_new_attack}) previously undisclosed malicious incidents involving over \$1.5M in value. In terms of efficiency, our system supports transaction-level parallelism and meets the real-time requirements of most mainstream blockchains.

\PP{Contributions}
We make the following key contributions:

\noindent $\bullet$  We present the first reactive detection generalization system that translates a single observed exploit into reusable detection for future imitative attacks. 

\noindent $\bullet$  We introduce a semantic extraction and logic matching framework that mitigates false positives caused by obfuscated benign transactions and improves resilience against evasive attack patterns, enabling high-fidelity detection.

\noindent $\bullet$  We implement a high-speed, high-precision detection pipeline that supports real-time analysis, meeting practical latency requirements for on-chain monitoring.

\noindent $\bullet$  Our system successfully uncovers 56 previously unrevealed attacks in real-world transaction datasets, demonstrating its practical utility and coverage beyond known exploits.

\section{Background and Motivation}

\subsection{Blockchain and DeFi}

Blockchain is a decentralized ledger that allows immutable and transparent record-keeping across a distributed network. Among its most influential innovations is the smart contract, a self-executing program deployed on the blockchain that enables programmable and automated interactions. Prominent platforms such as Ethereum~\cite{etherscan} and Solana~\cite{solana} provide robust infrastructures for smart contract execution, which serve as a foundational layer for decentralized financial applications (DeFi). These contracts operate autonomously once deployed, managing assets and logic without further intervention--making them powerful yet vulnerable to logic flaws or misuse.
DeFi refers to a suite of financial applications built on top of blockchains, aiming to provide open, permissionless alternatives to traditional financial services~\cite{mabruri2024dynamic}. From lending protocols and decentralized exchanges (DEXs)~\cite{hagele2024DEX} to synthetic assets and derivatives~\cite{schar2021Derivatives}, DeFi platforms have rapidly grown in complexity and adoption. With billions of dollars~\cite{Cymetrics} locked in DeFi contracts, the ecosystem has become an attractive target for adversaries. Unlike traditional systems, where centralized oversight might limit damage or provide rollback mechanisms, DeFi attacks are typically fast, irreversible, and often replicated across chains--exposing a critical need for timely and generalizable vulnerability detection.

\subsection{DeFi Attacks and Existing Detections}
DeFi attacks fall broadly into two categories: price-driven~\cite{xi2024pomabuster, mo2023priceFeed} and non-price~\cite{alhaidari2025nonPrice} attacks. Price-driven attacks manipulate pricing mechanisms, such as oracles or liquidity pools, to buy low and sell high, extracting universal tokens (e.g., ETH~\cite{katsiampa2019volatility}, stable-coins) at favorable rates. These often involve flash loans or arbitrage and depend heavily on capital. Tools like POMABuster~\cite{xi2024pomabuster} and DeFiRanger~\cite{wu2023defiranger} are tailored to detect such attacks by identifying abnormal value flows or predefined trading patterns, but their reliance on price-based signals makes them ill-suited for other exploit types.

Non-price attacks, on the other hand, exploit flaws in smart contract logic or protocol design, such as reentrancy~\cite{ji2021reentrancy}, accounting errors~\cite{zhang2024accounting}, or privilege misconfigurations~\cite{wu2023privilege}, without requiring capital. These exploits are diverse, subtle, and often obfuscated, making detection via hardcoded heuristics fragile. Tools like TxSpector~\cite{zhang2020txspector} propose handcrafted rules for specific patterns (e.g., unchecked calls or suicidal contracts) but lack breadth. Forta~\cite{Forta}, though proactive, relies on bytecode-level features that are sensitive to compiler noise, coding style, and obfuscation, and fail to generalize. While FlashGuard~\cite{alhaidari2025nonPrice} recognizes the importance of non-price exploits, it focuses on post-incident mitigation rather than timely detection. To date, no existing tool offers a robust and scalable solution for semantically detecting diverse non-price attacks, highlighting the need for generalizable approaches like ours that reason over attacker intent at the trace level.

\subsection{Motivation Example}\label{sec:back_motiv}

\begin{figure}[ht]
    \centering
    \includegraphics[width=0.98\linewidth]{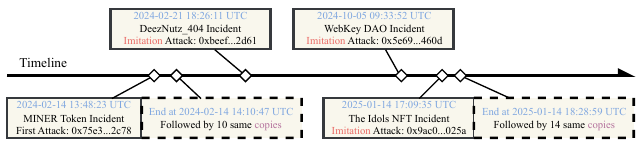}
    \Description{MINER Token Incident and Imitative Attack Cascade.}
    \caption{MINER Token Incident and Imitative Attack Cascade. \textit{Dashed boxes indicate repeated attacks from the same address ("Copies"), while "Imitation" refers to similar logic used by different addresses.}}
    \label{fig:motiv_flow}
\end{figure}






Recent DeFi incidents reveal a recurring pattern: once an exploit is discovered, it often leads to a series of \emph{imitative attacks} over time. While industry tools~\cite{gogol2024sok} sometimes detect the initial attack using ad hoc signals--such as exploiter addresses, Tornado~\cite{nadler2023tornado} activity, or large transfers--these detections are often incidental. Without understanding the underlying exploit logic, they struggle to capture subsequent variants. To better illustrate the limitations of current detection methodologies, we present a representative example in \autoref{fig:motiv_flow}--a non-price-driven exploit that occurred on February 14, 2024. The attacker leveraged a subtle business logic flaw: the contract failed to prohibit self-transfers, and a temporary balance variable was mistakenly used for withdrawal validation. This allowed the attacker to withdraw more funds than permitted without relying on any price manipulation or initial capital~\cite{MinerToken}. Alarmingly, even after the incident was publicly disclosed, we continued to observe similar attacks exploiting nearly identical flaws well into 2025~\cite{recur_idol_nft}. This long tail of incidents demonstrates the failure of existing reactive systems to generalize detections across such \emph{imitative attacks cascade}. From the Phalcon~\cite{phalcon} and DefiHackLab~\cite{defihacklab} incident reports, we find that over 750 attacks in the past three years are repetitions of previously disclosed exploits, with more than 37\% of the initial attacks having spawned 2-10 imitative copies. The cumulative loss attributed to these \emph{imitative attacks} amounts to \$2 billion.

\section{Problem Definition and Overview}\label{sec:ProblemScopeAndOverview}



\subsection{Problem Definition}\label{subsec:problemScope}
The observations above (\autoref{sec:back_motiv}) motivate our research goal: to \textbf{generalize rapid, reactive detection for \emph{imitative attack cascades}} once an initial attack instance is identified. To specify the problem, we introduce two concepts, \emph{Unobserved Attack} (\autoref{def:Unobserved}) and \emph{Reactive Detection Generalization} (\autoref{def:detection_generation}), that formalize our system's design goal.

Considering the case of the motivation example in \autoref{fig:motiv_flow}, where the initial exploit goes undetected by existing tools due to its novel logic. We define such an instance as an \emph{Unobserved Attack}:
\begin{definition}[Unobserved Attack]\label{def:Unobserved}
Let $\mathcal{T}$ denote the universe of transactions, $\mathcal{A} \subseteq \mathcal{T}$ the set of attack transactions, $\mathcal{N} = \mathcal{T} \setminus \mathcal{A}$ the set of benign transactions, and $\mathcal{P} = \{p_1, p_2, \dots, p_n\}$ the set of deterministic detection patterns predefined where $p_i: \mathcal{T} \to \{True, False\}$.
We define the set of \emph{Observed Attacks} under pattern set $\mathcal{P}$ as:
\[
\mathcal{A}_{\text{ob}}(\mathcal{P}) = \left\{ t \in \mathcal{A} \,\middle|\,
\begin{aligned}
&\exists p_i \in \mathcal{P},\ p_i(t) = \text{True} \\
&\text{and} \ \forall n \in \mathcal{N},\ p_i(n) = \text{False}
\end{aligned}
\right\}
\]
 Then, the set of \emph{Unobserved Attacks} is:
\[
\mathcal{A}_{\text{unob}}(\mathcal{P}) = \mathcal{A} \setminus \mathcal{A}_{\text{ob}}(\mathcal{P})
\]
\end{definition}
We say that a transaction $t$ is an \emph{Unobserved Attack} (with respect to $\mathcal{P}$) if $t \in \mathcal{A}_{\text{unob}}(\mathcal{P})$.
In other words, $t$ is an attack transaction that fails to be matched by any known detection rules implemented in current tools. To detect the \emph{Unobserved Attacks} to prevent subsequent damage, our goal is to \emph{Generalize Reactive Detection} for the \emph{Unobserved Attack}:
\begin{definition}[Reactive Detection Generalization]\label{def:detection_generation}
Given an \emph{Unobserved Attack} transaction $t \in \mathcal{A}_{\text{unob}}(\mathcal{P})$ with respect to a pattern set $\mathcal{P}$, a \emph{Reactive Detection Generalization} procedure constructs a new pattern $p^*$ such that:
\[
\forall n \in \mathcal{N}, p^*(t) = \text{True} \wedge p^*(n) = \text{False}
\]
Consequently, the transaction $t$ becomes detectable under the updated pattern set $\mathcal{P}' =  \mathcal{P} \cup \{p^*\}$, i.e., $t \in \mathcal{A}_{\text{ob}}(\mathcal{P}')$.
\end{definition}

The definitions above describe idealized pattern sets $\mathcal{P}$ and $\mathcal{P}'$ under the assumption of perfect precision. In practical implementations, this requirement is relaxed to tolerate minor false positives, provided that the overall detection performance remains acceptable. Nevertheless, achieving \emph{Reactive Detection Generalization} remains challenging for security experts due to (C1) the difficulty of abstracting diverse function semantics and (C2) matching logic in noisy, obfuscated traces. Manual analysis delays response and gives attackers further opportunities to act, highlighting the need for automated solutions to accelerate reaction time.

\subsection{Overview of GenDetect}\label{subsec:overview}
\begin{figure}[ht]
    \centering
    \includegraphics[width=0.90\linewidth]{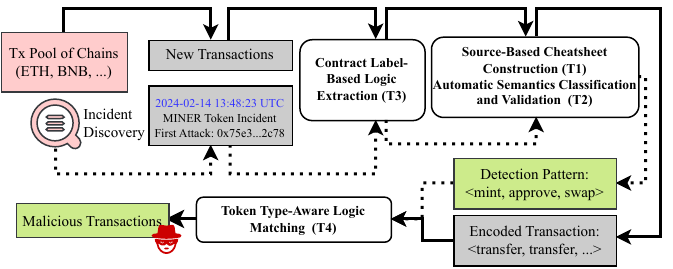}
    \Description{Overview diagram of GenDetect}
    \caption{The Overview of GenDetect}
    \label{fig:implementation}
\end{figure}

\PP{Overall Design}
With our goal formally defined, we now outline the overall design of \emph{GenDetect}, which operationalizes the path from challenge to insight to implementation. As shown in \autoref{fig:implementation}, the system takes two inputs: \circled{1} a newly discovered attack incident, which serves as a starting point for \emph{reactive detection generalization}, and \circled{2} a pool of pending transactions to be analyzed. To address the challenges introduced earlier, GenDetect adopts two key insights: \textbf{Source-based Semantic Extraction} and \textbf{Contract Label-based Logic Matching}. These insights are instantiated through four core components: I) Technique 1 and 2 implement Insight I by transforming raw transaction traces into semantically meaningful representations via contract source code analysis; II) Technique 3 and 4 realize Insight II by refining the trace into an abstracted logic path and matching it against the pattern.

\PP{Solving the Challenges}
The first challenge stems from the difficulty in accurately inferring the semantics of DeFi-specific function signatures (C1). Existing methods typically rely on word-level embeddings to classify function names. This works in some cases, e.g., \texttt{swapExactTokens}, \texttt{swap}, and \texttt{transferFrom} can be reasonably grouped with \texttt{swap} and\texttt{ transfer} due to shared intent. However, in other cases, this approach clearly fails, e.g., \texttt{deposit} is different from \texttt{beforeDeposit}. Despite similar word-level tokens, these functions often have entirely different roles: \texttt{deposit} triggers a financial operation, whereas \texttt{beforeDeposit} merely performs access checks. Such semantic misclassification undermines the reliability of transaction-level analysis.

We tackle this issue by anchoring our solution in a simple yet powerful insight: only source code can faithfully reflect a function's actual behavior. Fortunately, most valuable DeFi projects are open source. We capitalize on this property with a two-stage design. First, we propose \textbf{T1: Source-Based Cheatsheet Construction (\autoref{sec:Cheatsheet} \autoref{fig:Cheatsheet})}. We first use CodeBERT to roughly cluster function signatures in popular DeFi codebases then manually inspect them by real-world functionality. Each signature is annotated based on its corresponding financial intent (e.g., swap, transfer, borrow), forming a curated reference table. This cheatsheet enables accurate, human-aligned classification for all known signatures. However, this alone is insufficient for full coverage. As new contracts accompanied with new function signatures are constantly emerging. To support unseen signatures, we introduce \textbf{T2: Automatic Semantics Classification and Validation (\autoref{sec:Cheatsheet} \autoref{fig:HybridSemantics})}, which automatically fetches open-source code corresponding to new functions, uses CodeBERT to derive the most similar function stored in codebase, and then leverages LLM-based reasoning to validate the derived label. This hybrid strategy ensures our system remains robust and up-to-date, extending semantic understanding to novel cases in the wild.

The second challenge (C2) stems from the inherent complexity of transaction logic in DeFi attacks. Unlike malware detection or contract vulnerability analysis, there exists no standard approach for extracting and comparing the semantic intent of transaction traces. Existing tools often fall back on naive pattern matching techniques such as Longest Common Subsequence (LCS). However, naive LCS suffers from two key limitations: (i) It treats all function calls equally, failing to isolate security-critical actions from irrelevant or benign operations. As a result, traces of long, legitimate transactions dilute the attack signal, leading to false positives, and (ii) LCS is highly sensitive to the execution order. Many malicious transactions may follow the same logical steps as known attacks but in a different order, which reduces similarity scores and increases false negatives.

To address this, we observe that the essence of an attack is often concentrated in a small subset of operations initiated by the attacker and targeting two typical categories of DeFi protocols. Therefore, the solution lies in selectively extracting meaningful logic and applying similarity analysis that is robust to noise. We realize this insight through two steps. \textbf{T3: Contract Label-based Logic Extraction (\autoref{sec:LogicExtraction} \autoref{fig:Label-basedLogic})}. Leveraging publicly available Contract Labeling datasets (Phalcon and 4Bytes) and transaction metadata, we identify key participants such as DeFi protocol addresses and attacker-controlled contracts. We then prune the trace by filtering out internal or routine contract invocations and collapsing nested layers into a flattened trace that preserves only the essential operations. This drastically reduces irrelevant noise and improves the signal-to-noise ratio for subsequent analysis. \textbf{T4: Token Type-aware Logic Matching (\autoref{sec:LogicMatchin} \autoref{fig:TypeAwareSimilarity})}. The extracted logic is further partitioned based on the target token's economic role, distinguishing between core asset tokens (e.g., WETH, USDC) and protocol-specific tokens (e.g., DOGE, AAVE). These represent different attack intents: value extraction versus exploitation. For each token category, we compute the "Asymmetrical Normalized Set Difference (ANSD)" between traces, then aggregate weighted scores. This strategy boosts robustness in two ways: (1) benign traces tend to focus on a single category, while attacks involve both, reducing false positives; (2) using asymmetrical set-based rather than sequence-based comparison mitigates false negatives caused by minor reordering and inserted noise.

\section{The GenDetect}\label{sec:design}
Having introduced our two core challenges and the key insights that guided our solution design, we now turn to the concrete technical realization of these four techniques.
\subsection{Source-based Semantic Extraction}\label{sec:Cheatsheet}
\PP{Function Signature Cheatsheet Implementation}
To construct our Function Signature Cheatsheet, we begin by collecting historical attack data from DeFiHackLab~\cite{defihacklab} and Phalcon~\cite{phalcon}, extracting all function signatures involved in security incidents over the past five years. This process yields 1,272 unique function signatures linked to DeFi protocol projects.
Next, using the "Decoded Projects" dataset provided by Dune~\cite{dune}, we successfully indexed the corresponding implementation code for each of these functions. This allows us to build a comprehensive source code database.

\begin{figure}[ht]
    \centering
    \includegraphics[width=0.98\linewidth]{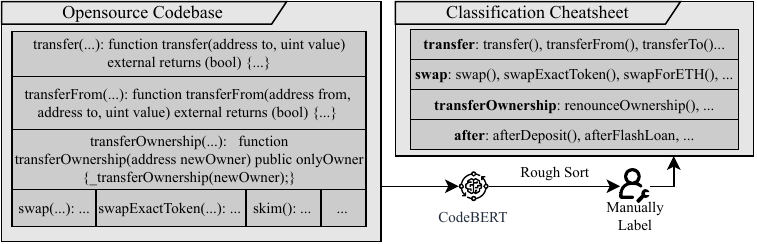}
    \Description{Cheatsheet Preparation}
    \caption{Function Signature Cheatsheet Construction (T1).\textit{The left is the CodeBase, a mapping from function signatures to their implementation code.
	The right is the final Function Cheatsheet, where each category includes multiple semantically similar function signatures. Importantly, clustering is based on source code, not just naming conventions.}}
    \label{fig:Cheatsheet}
\end{figure}

We then apply CodeBERT~\cite{codebert} to compute pairwise code-level semantic similarities across all collected function source code. For clustering, we adopt a K-means-based approach with an initial setting of $K = 80$, chosen heuristically based on functional diversity. The computation took approximately 3 hours.
Following clustering, we manually validated and refined the results. The final cheatsheet was expanded to contain 122 distinct function categories. Among them: \circled{1} 60 categories correspond to high-frequency financial or verification-related functions, which appeared at least twice across different attacks. These include core operations such as \texttt{transfer}, \texttt{swap}, \texttt{mint}, as well as verification hooks like \texttt{beforeDeposit} and \texttt{afterBorrow}. \circled{2} The remaining 62 categories represent low-frequency or project-specific functions, each called fewer than twice across five years of attack data. These include niche operations such as \texttt{emergencyBurn}, \texttt{burnAlp}, or \texttt{airDropReward}. This cheatsheet enables us to instantly map a known signature to its semantic category whenever encountered in future traces, significantly improving trace abstraction efficiency. \autoref{fig:Cheatsheet} provides a brief illustration of the process.

\begin{figure}[ht]
    \centering
    \includegraphics[width=0.90\linewidth]{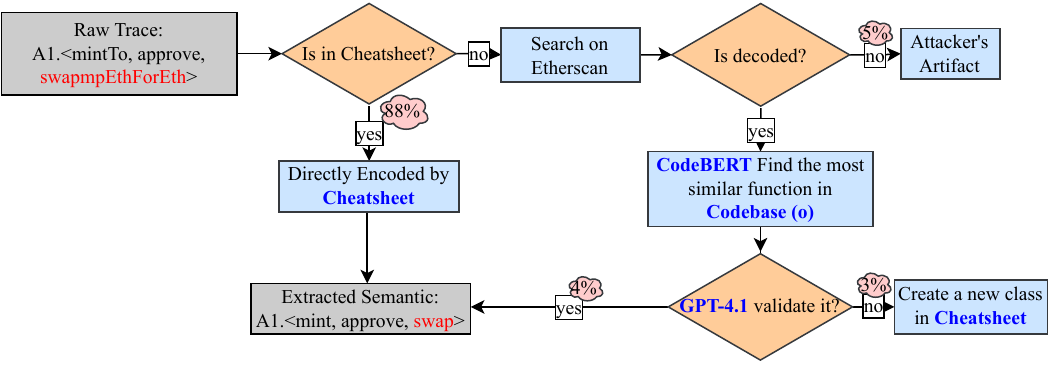}
    \Description{Semantic Extraction}
    \caption{Automatic Semantic Classification and Validation (T2). \textit{The red cloud icons indicates the relative proportion of data processed along each branch of the pipeline.}}
    \label{fig:HybridSemantics}
\end{figure}

\PP{Automatic Pipeline Implementation}
To handle unseen function signatures that are not covered by our cheatsheet, we introduce an Automatic Semantics Classification and Validation pipeline, as illustrated in \autoref{fig:HybridSemantics}. For any signature already present in the cheatsheet, we can directly retrieve its semantic label. For those not covered, we follow a three-step fallback mechanism: \circled{1} Contract Decoding Check: We query Etherscan API~\cite{etherscan} to check if the contract containing the unknown function has been decoded. If not, we discard it, as these are likely attacker-crafted and will be filtered out during logic extraction (see \autoref{sec:LogicExtraction} \autoref{fig:Label-basedLogic}). \circled{2} Code Similarity Search: If decoded, we retrieve its implementation code and compare it against a reduced reference codebase $\text{CodeBase}^{(o)}$, which is a curated subset of representative implementations from each cheatsheet category. This optimization significantly reduces the computational cost of CodeBERT-based similarity matching. \circled{3} Validation via GPT-4.1: Once the most similar category is identified, we perform a final heuristic validation using GPT-4.1. If GPT confirms the categorization, the function is assigned to that category; otherwise, we create a new class in the cheatsheet for future reference. This layered fallback ensures robust semantic coverage, even for novel or rarely used DeFi functions that are not part of the initial cheatsheet database.

\subsection{Contract Label-based Logic Extraction}\label{sec:LogicExtraction}

\begin{figure}[ht]
    \centering
    \includegraphics[width=0.98\linewidth]{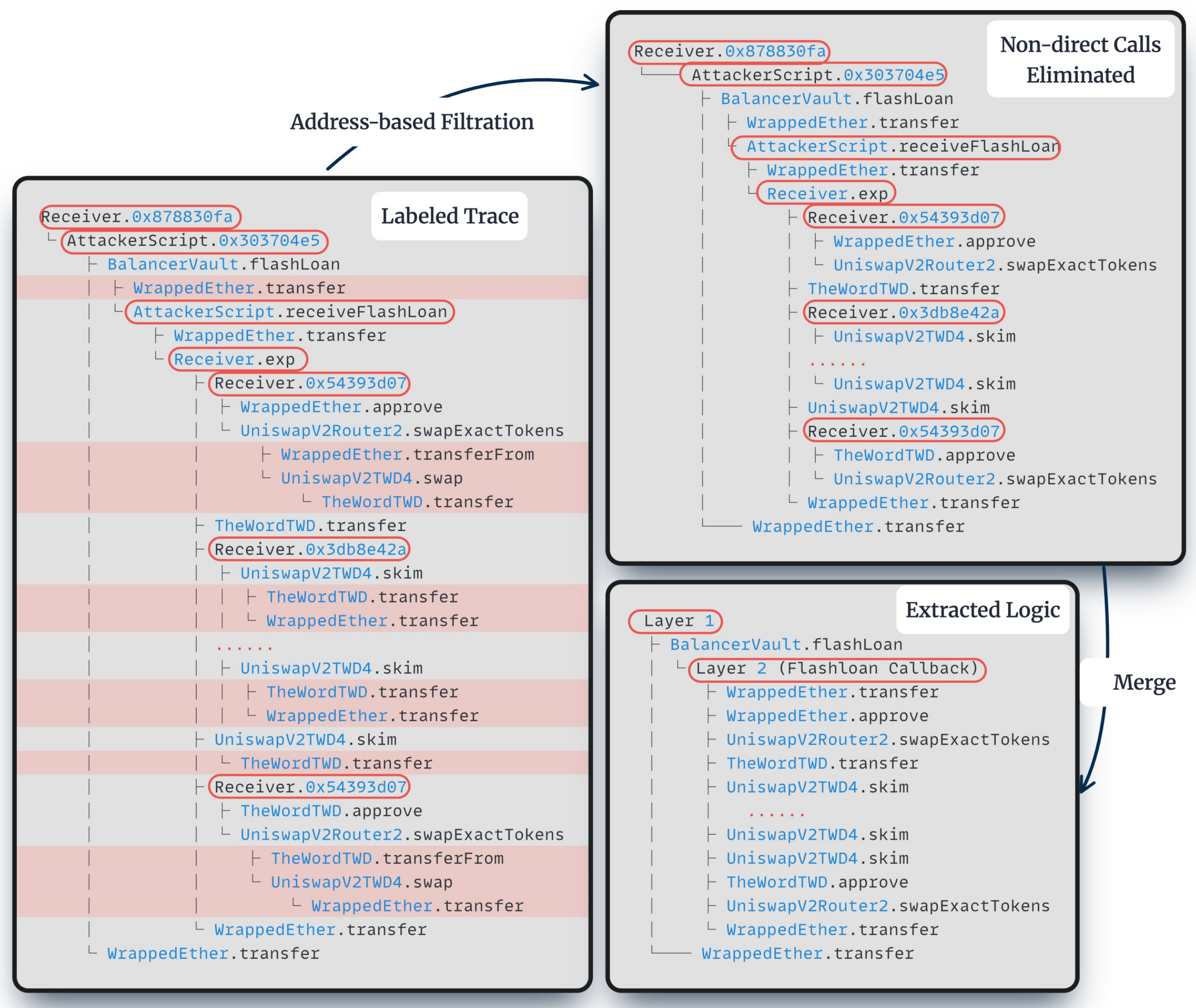}
    \Description{Logic Extraction Snippet}
    \caption{Demonstration of Address-based Logic Extraction (T3).}
    \label{fig:Label-basedLogic}
\end{figure}

\PP{Address Labelling} We illustrate the logic extraction process in \autoref{fig:Label-basedLogic}. The first step involves tagging each address invocation in the trace. Notably, most well-known or high-value project addresses are labeled in the 4Bytes~\cite{4bytes_contract} database. For less mainstream or newer project addresses, we rely on labels provided by the Phalcon API~\cite{phalcon}. While we cannot fully eliminate the possibility that some recently deployed projects may not yet be included in these datasets, we note that our system's accuracy can be further improved as the labeling coverage of community-maintained datasets increases. Addresses that lack labels, or are explicitly marked as exploiter, or are directly called by the transaction sender, are grouped into the category \texttt{AttackerScript}. These will serve as the primary focus for the next phase of logic extraction.

\PP{Filtration} In the second step, we retain only direct invocations from the sender or identified \texttt{AttackerScript}, as only these invocations reflect the attacker's intentional actions. Downstream calls that originate from the protocol itself (e.g., recursive contract calls or triggered hooks) are excluded because they represent protocol logic rather than attacker intent. Finally, we perform layer restructuring on the remaining invocations. Specifically, for any invocation made by \texttt{sender.call} or \texttt{AttackerScript.call} we remove the wrapper call itself and lift its internal subcalls one level up in the hierarchy. This adjustment reflects the fact that the outer attacker-generated calls are often meaningless or not decoded, whereas the meaningful behavior is embedded in the downstream interactions with protocol contracts. By eliminating these superficial wrappers while preserving the original call depth of the subcalls, we obtain a cleaner and more interpretable Extracted Logic that better captures the attacker's true intent.

\subsection{Token Type-Aware Logic Matching}\label{sec:LogicMatchin}

\begin{figure}[ht]
    \centering
    \includegraphics[width=0.90\linewidth]{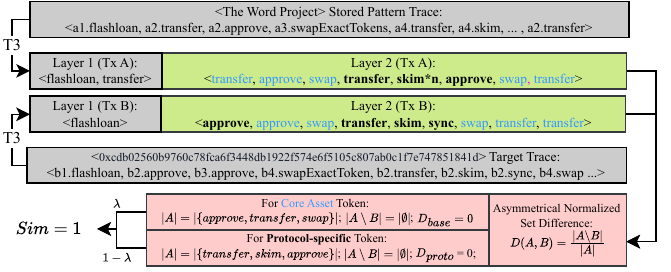}
    \Description{Similarity Analysis}
    \caption{Demonstration of Type-aware Logic Similarity Matching (T4).}
    \label{fig:TypeAwareSimilarity}
\end{figure}

\PP{Type Labelling} This module enhances the accuracy of logic similarity by introducing token type awareness into the matching process. Specifically, we classify each invocation in the Extracted Logic (output of \autoref{sec:LogicExtraction}) into two categories based on the token involved: \circled{1} Core Asset Operations: Interactions with widely-used tokens such as ETH, BNB, USDT, etc. \circled{2} Protocol-specific Operations: Interactions with tokens that are unique to a particular protocol or newly deployed projects, such as DOGE derivatives or bespoke tokens like RisyToken. As illustrated in the green box of \autoref{fig:TypeAwareSimilarity}, different token types are marked with distinct colors, e.g., blue for core assets and black for protocol-specific assets. Next, as shown in the red box, given two transactions A and B, we compute two separate similarity scores: (i) $\text{Sim}_{\text{core}}(A, B)$: similarity between core asset operations in A and B and (ii) $\text{Sim}_{\text{proto}}(A, B)$: similarity between protocol-specific operations in A and B.
The final similarity score is then computed as a weighted average of the two: $\text{Sim}_{\text{final}}(A, B) = \lambda \cdot \text{Sim}_{\text{core}} + (1-\lambda) \cdot \text{Sim}_{\text{proto}}$

\begin{figure}[ht]
    \centering
    \includegraphics[width=0.90\linewidth]{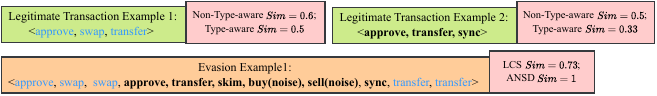}
    \Description{Similarity Analysis for Examples of Legitimate Transactions and Evasive Transactions}
    \caption{Benign vs. Evasive Transactions: Similarity Analysis}
    \label{fig:SimilarityExample}
\end{figure}

This token-type-aware decomposition improves the contrast between benign and malicious behavior. Benign transactions tend to contain only one category of operations, while imitative attacks often span both, making this split particularly effective at reducing false positives (FP). An illustrative example demonstrating the benefit of this separation will be presented in the case study (\autoref{fig:SimilarityExample}).

\PP{Asymmetrical Normalized Set Difference}
To compute the similarity between two extracted logic sequences, we adopt a set-based approach rather than a sequence-based one like LCS (Longest Common Subsequence). This choice is motivated by the observation that imitative attacks frequently involve logic mutation: while the core behaviors remain similar, their invocation order and auxiliary operations often differ. Sequence-based methods like LCS are sensitive to such ordering variations and thus perform poorly in this context. In contrast, set-based methods are more robust to such transformations.

We use a metric called Asymmetrical Normalized Set Difference (ANSD) for similarity calculation. Given two transactions $A$ and $B$, we define the similarity from $B$ to $A$ as: $\text{Sim}(A, B) = 1 - \frac{|A \setminus B|}{|A|}$. This expression measures how much of the reference logic $A$ (which is a confirmed attack) is preserved in $B$. Based on this metric, our detection rule is straightforward: if the final similarity score $\text{Sim}_{\text{final}}(A, B) \geq \tau$ ($\tau$ is a predefined threshold), we flag transaction $B$ as an imitative attack.

\PP{Rationale for ANSD Design}
The rationale for the asymmetric design is twofold: 

\textit{(i) Why is the denominator $|A|$, rather than $|B|, |A \cup B|$?} Since $A$ is a verified attack pattern, it serves as the ground truth. Our goal is to assess whether $B$ contains all the necessary elements to reproduce the same exploit. $A$ denominator of $|A|$ ensures that the metric reflects the proportion of the attack pattern retained.

\textit{(ii) Why use $A \setminus B$ as the numerator?} We assume that attack pattern $A$ is sufficient but may not minimal: any transaction that omits elements from $A$ is less likely to constitute the same threat. Hence, the more elements from $A$ that are missing in $B$, the lower the similarity. In contrast, using $B \setminus A$ would penalize irrelevant additions, which may be benign noise rather than part of the exploit, and is therefore less meaningful in this context.

By using the confirmed attack $A$ as the reference, the metric disregards any additional noise or evasion operations introduced in $B$, ensuring that only the omission of logic from $A$ leads to a lower similarity score. This design makes evasion behaviors, such as adding misleading calls, largely ineffective, reducing false negatives (FN). A example illustrating this robustness is provided in \autoref{fig:SimilarityExample}.

\PP{Case Study: Filtering Benign Transactions and Tolerating Mutations}
We present three illustrative cases in \autoref{fig:SimilarityExample} to highlight the advantages of type-aware matching and our Asymmetrical Normalized Set Difference (ANSD) metric. The first two in green boxes are benign transactions: \circled{1} on core assets; \circled{2} on protocol-specific assets. In both cases, the non-type-aware method yields inflated similarity scores, while the type-aware variant correctly down-weights them, based on our insight that transactions involving only one type of token operation are unlikely to reproduce the full logic of an attack. The third example is an evasion attempt: \circled{3} it mutates call order (\texttt{approve}) and adds irrelevant noise (\texttt{buy}, \texttt{sell}). Traditional sequence-based methods like LCS are sensitive to such changes and produce low similarity, whereas our ANSD remains stable by focusing on functional coverage and ignoring order and noise. These cases validate that type-aware ANSD improves discriminative power, suppressing false positives and maintaining robustness under adversarial mutations.

\begin{figure}[ht]
    \centering
    \includegraphics[width=0.9\linewidth]{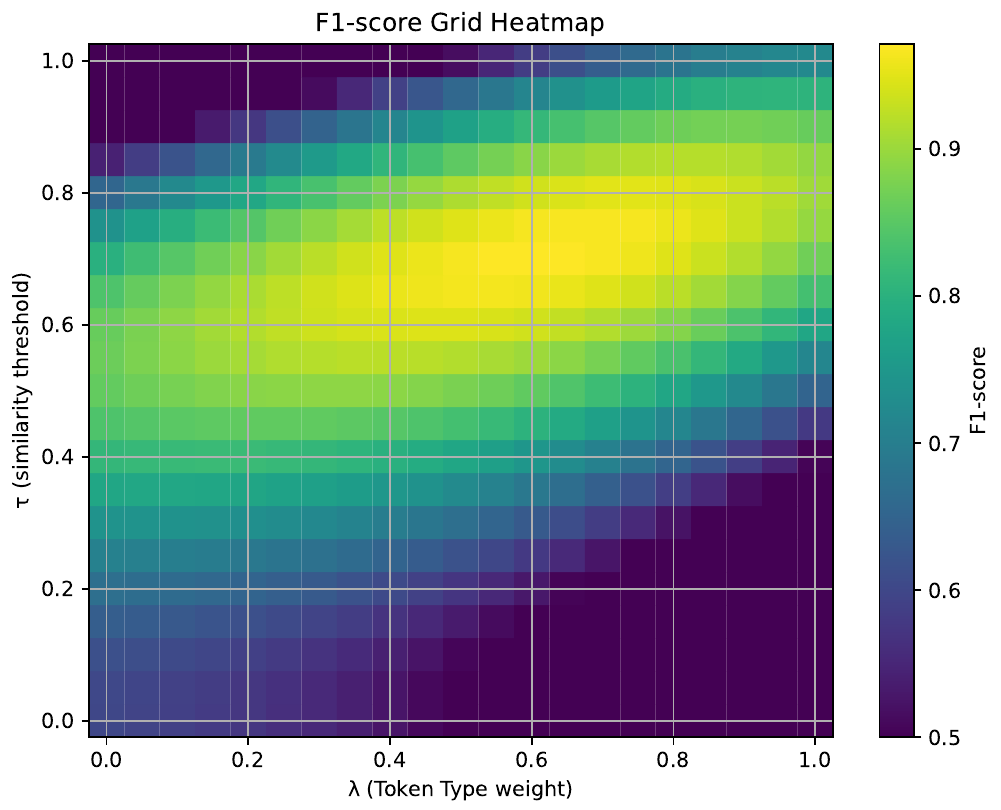}
    \Description{Hyperparameter($[\lambda, \tau]$) Optimization through Nested 4-fold Cross-validation}
    \caption{Hyperparameter($[\lambda, \tau]$) Optimization through Nested 4-fold Cross-validation}
    \label{fig:hyperparameter}
\end{figure}

\PP{Hyperparameter Optimization for Logic Matching}
The overall logic matching predicate is defined as: \[
p_{A}{(B)} =
\begin{cases}
\text{True}, & \text{if } \lambda \cdot \text{Sim}_{\text{core}} + (1 - \lambda) \cdot \text{Sim}_{\text{proto}} \geq \tau \\
\text{False}, & \text{otherwise}
\end{cases}
\],
where $\lambda \in [0, 1]$ adjusts the contribution between core asset type and protocol-specific type operation, and $\tau \in [0, 1]$ adjusts the final similarity threshold to determine whether a transaction is flagged as malicious. 
We tune both hyperparameters ($\lambda$, $\tau$) using grid search on a held-out validation split within each fold during nested 4-fold cross-validation (\autoref{fig:hyperparameter}). Specifically, for each of the four folds, we split the training portion (801 samples from 534 attack and 534 benign examples) into a "9:1" internal split: approximately 720 transactions for training and 81 for validation. The hyperparameters are selected based on the F1-score performance on this internal validation set. This strategy ensures that the outer test folds remain completely untouched during tuning, preserving the integrity and fairness of cross-validation evaluation. The final parameter ranges ($\lambda \in [0.52, 0.66], \tau \in [0.68, 0.72]$) correspond to the high-performing region highlighted in \autoref{fig:hyperparameter}.

\section{Evaluation}\label{sec:experiment}

We evaluate GenDetect by answering the following questions:

\vspace{-\parskip} \noindent \textbf{RQ1:} How accurate is our detection framework, and how does each module contribute to its performance?

\vspace{-\parskip} \noindent \textbf{RQ2:} Can our system discover new, previously undetected attacks in the wild?

\vspace{-\parskip} \noindent \textbf{RQ3:} Is our system capable of operating under real-time constraints, and what is the computational cost of each core module?


\PP{Baseline Tools}
We evaluate our system against four representative detection tools: TxSpector~\cite{zhang2020txspector}, POMABuster~\cite{xi2024pomabuster}, DeFiRanger~\cite{wu2023defiranger}, and Forta~\cite{Forta}. TxSpector detects low-level logic bugs (e.g., reentrancy, suicidal contracts, unchecked calls) using hand-crafted rules on decoded traces; we adopt the official implementation~\cite{txSpector_code}. POMABuster targets financial exploits via pattern matching on funding flows, and we directly use its released code~\cite{POMABuster_code}. DeFiRanger extends this pattern-based strategy with additional rules, but only an unofficial implementation exists--released by the authors of POMABuster~\cite{POMABuster_code}--which covers only price-related patterns. To ensure a fair comparison, we supplement this baseline with our implementation of the missing non-price rules based on the original paper. Forta is a commercial monitoring framework based on contract bytecode; we use its open-source agent code~\cite{Forta_code} and retrain it on our benchmark. Unlike Forta, which requires training on bytecode, the other tools operate with fixed rules. Our system is trained on the same benchmark dataset as Forta for detection generalization and hyperparameter optimization.


\PP{Experimental Environment}\label{sec:exp_setup}
All experiments were conducted on a Linux server running Ubuntu 24.04 LTS, equipped with an Intel Xeon Platinum 8268 CPU (96 logical cores at 2.90GHz) and 256 GB of RAM. Our implementation is developed in Python 3.11 and relies on the following major libraries: OpenAI API (openai 1.58.1), code-bert-score 0.4.1 and scikit-learn 1.5.1. All model inference and detection tasks were executed on the CPU only.

\subsection{RQ1: Detection Accuracy}\label{sec:exp_rq1_acc}

This section evaluates the detection accuracy of our system and baseline tools under both in-distribution and out-of-distribution (OOD) settings. We first report the cross-validation performance of all compared methods on the benchmark dataset, including the results of our ablation variants (\autoref{tab:cv-combined}). Then we evaluate it under skewed ratios and mixed category settings (\autoref{tab:cv-combined-mixed}). Next, we evaluate the zero-shot generalization ability of each system using the held-out dataset. To avoid potential data leakage and ensure a fair evaluation, we perform cross-domain evaluation by swapping the training and test sources, i.e., using DeFiHackLab for training and Phalcon for testing, and vice versa (\autoref{tab:detection-coverage-independent}). These evaluation setups allow us to assess GenDetect's detection rate across real-world DeFi attack scenarios.

\begin{table}[ht]
\scriptsize
\centering
\caption{4-Fold Cross-Validation Results (Mean $\pm$ Std) Across Detection Methods and Ablations on DEX-Only Dataset. \textit{w/o = without, SX = Semantics Extraction, LX = Logic Extraction, TTA = Token Type-aware, LCS/ANSD = replace Asymmetrical Normalized Set Difference with Longest Common Subsequence}}
\label{tab:cv-combined}
\begin{tabular}{lccccc}
\toprule
\textbf{Method} & \textbf{F1-score} & \textbf{FPR} & \textbf{FNR} & \textbf{Accuracy} & \textbf{Recall} \\
\midrule
\textbf{GenDetect}     & \textbf{.98 $\pm$ .01} & \textbf{.01 $\pm$ .01} & 
\textbf{.03 $\pm$ .01} & \textbf{.98 $\pm$ .01} & \textbf{.97 $\pm$ .01}\\
w/o SX  & .42 $\pm$ .15 & .44 $\pm$ .06 & .61 $\pm$ .15 & .48 $\pm$ .11 & .39 $\pm$ .15\\
w/o LX & .72 $\pm$ .02 & .22 $\pm$ .08 & .30 $\pm$ .06 & .74 $\pm$ .02 & .70 $\pm$ .06\\
w/o TTA   & .90 $\pm$ .02 & .17 $\pm$ .04 & .03 $\pm$ .01 & .89 $\pm$ .02 & .97 $\pm$ .01\\
LCS/ANSD  & .91 $\pm$ .03 & .01 $\pm$ .01 & .13 $\pm$ .05 & .92 $\pm$ .03 & .87 $\pm$ .05\\
\addlinespace
Forta   & .91 $\pm$ .02 & .16 $\pm$ .03 & .04 $\pm$ .02 & .90 $\pm$ .02 & .96 $\pm$ .02\\
DefiRanger  & .77 $\pm$ .06 & .23 $\pm$ .03 & .22 $\pm$ .09 & .77 $\pm$ .05 & .78 $\pm$ .09\\

\bottomrule
\end{tabular}
\end{table}

\begin{table}[ht]
\scriptsize
\centering
\caption{Cross-Validation Results of GenDetect under Skewed Malicious:Benign (M:B) Ratios on DEX-Only Dataset and Mixed Dataset including Aggregator, Staking, DEX, and NFT Transactions.}
\label{tab:cv-combined-mixed}
\begin{tabular}{lccccc}
\toprule
\textbf{M:B Ratio}& \textbf{F1-score(\%)} & \textbf{FPR(\%)} & \textbf{FNR(\%)} & \textbf{Accuracy(\%)} & \textbf{Recall(\%)}\\
\midrule
1:1 (DEX)  & 98.0 $\pm$ 1.1 & 0.9 $\pm$ 0.9 & 3.1 $\pm$ 1.4 & 98.0 $\pm$ 1.1 & 96.9 $\pm$ 1.4\\
1:5 (DEX)  & 97.6 $\pm$ 1.1 & 2.3 $\pm$ 1.2 & 2.3 $\pm$ 0.9 & 97.6 $\pm$ 1.1 & 97.6 $\pm$ 0.9\\
1:25 (DEX)  & 95.7 $\pm$ 1.4 & 3.8 $\pm$ 1.3 & 4.6 $\pm$ 1.6 & 95.8 $\pm$ 1.4 & 95.4 $\pm$ 1.6\\
1:1 (Mix)  & 98.6 $\pm$ 0.9 & 0.0 $\pm$ 0.0 & 2.6 $\pm$ 1.7 & 98.7 $\pm$ 0.8 & 97.3 $\pm$ 1.7\\
1:5 (Mix) & 98.3 $\pm$ 0.8 & 0.0 $\pm$ 0.0 & 2.9 $\pm$ 1.7 & 99.5 $\pm$ 0.2 & 97.1 $\pm$ 1.7\\
1:25 (Mix) & 97.9 $\pm$ 0.6 & 0.0 $\pm$ 0.0 & 3.7 $\pm$ 1.5 & 99.7 $\pm$ 0.0 & 96.2 $\pm$ 1.5\\
\bottomrule
\end{tabular}
\end{table}

\PP{Setup}
As a preliminary evaluation, we use 4-fold cross-validation to evaluate all methods on a balanced benchmark dataset consisting of 534 malicious traces from DeFiHackLab~\cite{defihacklab}, and 534 benign traces sampled from decentralized exchange (DEX) transactions recorded on the Dune platform~\cite{dune}. 
To supplement its performance under real-world data distributions, we further conduct evaluation on a mixed dataset containing 13,300 transactions from various DeFi categories. It proportionally includes DEX~\cite{dex_trades}, Aggregator~\cite{dex_aggregator_trades}, Staking ~\cite{staking_ethereum_flows}, and NFT~\cite{nft_trades} contracts.
Beyond the standard balanced (1:1) setting, we also evaluate the robustness of GenDetect under skewed malicious-to-benign (M:B) ratios of 1:5 and 1:25 on the mixed dataset, reflecting real-world scenario where benign transactions outnumber malicious ones.
For tools other than Forta, we directly use decoded transaction traces as input, as they are capable of processing features derived from execution traces. However, Forta only operates on deployed smart contract. To accommodate this input format, we extract the corresponding contract code for each relevant transaction from data sources and construct a parallel input set for Forta's evaluation and training.

In the \textbf{cross-validation setting}, we compare GenDetect with Forta and DeFiRanger. We exclude POMABuster and TxSpector due to their narrow scope: the former focuses solely on price oracle manipulation, while the latter targets a small set of low-level bugs like reentrancy and unchecked calls. Their limited coverage restricts their value in broader benchmark comparisons. In contrast, Forta and DeFiRanger capture higher-level behavioral patterns and support a wider range of attack types, making them more suitable for diverse evaluation.
In the \textbf{zero-shot setting} on the Phalcon and DeFiHackLab dataset~\cite{phalcon,defihacklab}, we conduct cross-domain evaluation, i.e., using DeFiHackLab for training and Phalcon for testing, and vice versa. This setup further extinguish the potential of data leakage caused by usage of address label (\autoref{sec:LogicExtraction}) provided by Phalcon.
We include all five tools: GenDetect, Forta, DeFiRanger, TxSpector, and POMABuster to assess performance on previously unseen attacks.
Notably, while GenDetect and Forta require training on the benchmark dataset, the detection rules of POMABuster, DeFiRanger, and TxSpector are fixed, illustrating a key distinction between trained and pattern-based systems.

\PP{Cross-Validation Results}
As shown in \autoref{tab:cv-combined}, GenDetect significantly outperforms both Forta and DeFiRanger across all major evaluation metrics on the cross-validation benchmark. Specifically, we achieve an F1-score of \textbf{0.98}, with a false positive rate (FPR) of only \textbf{0.01}. In contrast, Forta exhibits over \textbf{16$\times$} higher FPR, while DeFiRanger fails to capture a substantial portion of semantic attacks, resulting in an F1-score of just \textbf{0.63}. In \autoref{tab:cv-combined-mixed}, with ratio changing from 1:1 to 1:25, the F1-score decreases by 0.8\%, and FNR increases by 1\%. GenDetect maintains consistent and reliable performance even when benign transactions substantially outnumber malicious ones.

\emph{1) Understanding Ablation Study.}
To further understand the contribution of each design component. The \textbf{ablation results} show that removing any single module leads to a considerable drop in performance. Removing the semantics extraction module (w/o SX) leads to the most severe degradation, with the F1-score dropping over 50\% and the false negative rate (FNR) increasing to 0.61. This highlights the critical role of source-based semantic grouping in recognizing attack behaviors that would otherwise be misclassified or missed entirely. Excluding the logic extraction component (w/o LX) results in moderate performance degradation (F1 = 0.72), primarily due to increased noise from irrelevant invocations, which elevates both false positive and false negative rates. Disabling the token type-aware matching (w/o TTA) yields a relatively smaller drop (F1 = 0.90), with a noticeable increase in false positives, indicating its effectiveness in improving discrimination between benign and malicious traces. When replacing our set-based similarity metric with the sequence-sensitive LCS algorithm (LCS/ANSD), the FNR rises sharply to 0.13, confirming our hypothesis that LCS is vulnerable to minor mutations or reordering within transactions.

\emph{2) Impact of Dataset Categories.}
To examine the robustness of GenDetect across different categories of transactions, we evaluate its performance on a mixed-categories dataset (\autoref{tab:cv-combined-mixed}). The results show that FPR and FNR slightly decreased in the mixed setting compared to DEX-only setting. This improvement is due to the structural characteristics of the mixed dataset: unlike DEX protocols, which often involve long and complex multi-hop swap spanning multiple contracts, other application types such as NFT marketplaces typically feature shorter and more deterministic transaction patterns, e.g., single buy, sell, or approve calls. And staking systems mainly involve straightforward deposit or withdraw operations with limited contract depth. These short and semantically consistent transaction traces are easier to distinguish from malicious behaviors, thereby slightly lowering the overall error rates.

\emph{3) Impact of Skewed Data Ratio.}
\autoref{tab:cv-combined-mixed} shows GenDetect's performance on datasets with progressively skewed malicious-to-benign ratios. We observe that as the proportion of benign transactions increases, FPR and FNR rises slightly. Since the training process of hyperparameters is based on the benign-dominant distribution, it slightly increases the decision threshold $\tau$ (\autoref{sec:LogicMatchin}) for detecting malicious traces, leading to a few attacks being missed. Meanwhile, with more benign samples in the test set, the absolute number of false positives also increases marginally. But the overall effect remains minimal, and GenDetect maintains stable performance even under highly imbalanced conditions.

\emph{4) Understanding Baseline Tools' Limitations.}
Forta reports high accuracy on its benchmark~\cite{Forta}, but performs significantly worse in our setting. This discrepancy is primarily due to its detection paradigm: Forta relies on features extracted from attacker contracts themselves, assuming either source code availability or strong bytecode patterns. However, in real-world settings, most attacker contracts are unverified, obfuscated, or inaccessible, limiting the effectiveness of contract-level detection. In contrast, our method does not require visibility into the attacker's contract implementation. Instead, we analyze execution traces and focus on decoded, labeled protocol interactions to derive the attacker's intent. This trace-level semantic modeling allows us to generalize more effectively under realistic conditions with low attack contract transparency, explaining the superior performance of GenDetect on real-world attacks.
DeFiRanger, while incorporating richer patterns than low-level tools like TxSpector, still relies heavily on exact rule matching and fragile execution signatures. This leads to high false negative rates when faced with semantically similar but structurally variant attacks. Its limited generalization stems from overfitting to specific symbolic patterns, whereas our semantic encoding enables more robust behavioral abstraction, yielding superior coverage across diverse real-world exploits.

\begin{table}[ht!]
\centering
\caption{Summarized Cross-domain Evaluation: Zero-shot Detection Coverage on Phalcon or DeFiHackLab Attack Dataset. \textit{\checkmark = detected, $\triangle$ = running error, blank = not detected.} (For full results, please see \href{https://github.com/NobodyIsAnonymous/GenDetect_ICSE2026/blob/main/zero-shot_evaluation.pdf}{Github}~\cite{GenDetect_Zero-shot,GenDetect_Zero-shot_DeFiHackLab})}
\label{tab:detection-coverage-independent}
\resizebox{0.45\textwidth}{!}{
\begin{tabular}{lcccccccc}
\toprule
\textbf{Projects} & \multicolumn{4}{c}{\textbf{GenDetect (Ablation)}} & \multirow{2}{*}{\textbf{Forta}} & \multirow{2}{*}{\textbf{POMA}} & \multirow{2}{*}{\textbf{DeFiR}} & \multirow{2}{*}{\textbf{TxSpec}} \\
\cmidrule(lr){2-5}
From Phalcon~\cite{phalcon}  & \textbf{Full} & \textbf{w/o SX} & \textbf{w/o LX} & \textbf{w/o TTA} & & & & \\
\midrule
               MIM\_Spell & \checkmark &            &  \checkmark &            &            & $\triangle$ &             &            \\
                    Zoth  &            &            &  &            &            & $\triangle$ & $\triangle$ &            \\
         1inch Fusion V1  & \checkmark &            &  \checkmark &            & \checkmark &  \checkmark &  \checkmark &            \\
                  Infini  &            &            &             &            &            & $\triangle$ & $\triangle$ &            \\
            HegicOptions  & \checkmark & \checkmark &  \checkmark &            & \checkmark &  \checkmark &             &            \\
                   Bybit  &            &            &             &            &            & $\triangle$ & $\triangle$ &            \\
            \multicolumn{9}{l}{\textit{\small (Remaining rows omitted for brevity; full table at \href{https://github.com/NobodyIsAnonymous/GenDetect_ICSE2026/blob/main/zero-shot_evaluation.pdf}{Repo}~\cite{GenDetect_Zero-shot}.)}} \\
\midrule
\textbf{Coverage}  & \textbf{.7656} & .2813 & .5625 & .6094 & .6563 & .2813 & .4375 & .3125 \\
\bottomrule
\toprule
\multicolumn{2}{l}{From DeFiHackLab~\cite{defihacklab}} &  &  &  &  &  \\
\midrule
0vix & \checkmark &            &  \checkmark &  \checkmark  &  \checkmark & \checkmark &             &            \\
0x0DEX  &  \checkmark & \checkmark & \checkmark & \checkmark  & \checkmark & $\triangle$ & $\triangle$ &            \\
AES\_1  &  \checkmark  &            &  \checkmark   &  \checkmark  & \checkmark   & \checkmark & \checkmark &            \\
AES\_2  & \checkmark & \checkmark &  \checkmark &            & \checkmark &  \checkmark &             &            \\
AIS  &            &            &             &            &            &          &           &            \\
Allbridge  &  \checkmark  &  \checkmark & \checkmark  & \checkmark   &            & \checkmark & \checkmark &            \\
\multicolumn{9}{l}{\textit{\small (Remaining rows omitted for brevity; full table at \href{https://github.com/NobodyIsAnonymous/GenDetect_ICSE2026/blob/main/DeFiHackLab_zero_test.csv}{Repo}~\cite{GenDetect_Zero-shot_DeFiHackLab}.)}} \\
\midrule
\textbf{Coverage}  & \textbf{.7284} & .3824 & .5009 & .5697 & .6481 & .2485 & .3613 & .1950 \\

\bottomrule
\end{tabular}
}
\end{table}

\PP{Zero-Shot Results}
We evaluate the detection rate of all tools on held-out, real-world datasets consisting of 726 attack incidents reported by Phalcon~\cite{phalcon} and DeFiHackLab~\cite{defihacklab}. \autoref{tab:detection-coverage-independent} shows the summarized results (\href{https://github.com/NobodyIsAnonymous/GenDetect_ICSE2026/blob/main/zero-shot_evaluation.pdf}{Github Repo}~\cite{GenDetect_Zero-shot,GenDetect_Zero-shot_DeFiHackLab} for full results): GenDetect achieves a coverage rate of \textbf{77/73\%}, significantly outperforming Forta (66/65\%), DeFiRanger (44/36\%), TxSpector (31/20\%), and POMABuster (28/25\%). Among our ablation variants, the model without the semantic extraction module covers only \textbf{28\%-38\%} of incidents, indicating a nearly \textbf{50\%} drop in detection capacity. Removing either the logic extraction or token type-aware module reduces coverage to below 60\%, reflecting 16-22\% performance degradation.

\emph{1) Understanding Detection Rate Drop.}
As expected, detection performance in the zero-shot setting is lower than in the cross-validation experiments. This is largely due to the complete disjointness between the test set and the training set, which is a common source of generalization loss in real-world detection tasks. Nevertheless, our performance drop is moderate: compared to the detection rate (recall) reduction of 31-32\% observed in Forta, our model exhibits only a 21-25\% decline, suggesting better stability in the face of data in the wild. Upon closer inspection, we find that most of the undetected attacks fall into edge-case categories that lie outside the observable transaction semantics. In particular, several missed incidents involve \textbf{private key leakage}, where the exploit resembles a side-channel attack with no observable execution anomalies on-chain. Another group of hard-to-detect samples involves \textbf{unverified call input attacks}, where a single external call results in the entire asset drain, yet leaves behind little to no semantic trail. These cases are inherently difficult to capture with any trace-based or behavior-level approach, and we acknowledge them as current limitations of our design.


\emph{2) Understanding Baseline Tools' Limitations.}
Our model substantially outperforms all baseline tools under the zero-shot setting. Forta's performance is limited by its reliance on code-level features, which are sensitive to developer coding style and compiler optimizations and fail to robustly reflect attack intent. POMABuster, DeFiRanger, and TxSpector suffer from similar weaknesses: all are pattern-based systems with hardcoded heuristics, offering narrow coverage and low adaptability to novel attack variants. Furthermore, many transactions in the Phalcon dataset lack the specific signals required by these systems, e.g., POMABuster requires token price information that may be unavailable for unlisted assets, such as NFT-related transactions or those involving project-specific "internal token ledger" (i.e., with no publicly accessible price data)~\cite{recur_idol_nft}, leading to incomplete execution or detection failure.

\emph{3) Potential Data Leakage and Resolution.}
Since our method leverages manually annotated contract address labels from the Phalcon dataset, such label alignment may introduce a potential implicit advantage to the detection model, as both the labels and the transactions originate from the same source organization.
To rule out potential data leakage from shared label sources, we perform a cross-domain evaluation using the independent DeFiHackLab dataset. Although the coverage slightly decreases from 76\% to 73\% (\autoref{tab:detection-coverage-independent}), the overall performance remains consistent, indicating that any implicit linkage between datasets does not affect GenDetect's usability or the validity of our conclusions.

\subsection{RQ2: New Attack Discovery by GenDetect}\label{sec:exp_RQ2_new_attack}

In this section, we investigate the capability of our system to identify previously undisclosed attacks from real-world transaction data. We summarize the newly discovered attack instances in a comprehensive table on \href{https://github.com/NobodyIsAnonymous/GenDetect_ICSE2026/blob/main/discovered_attacks.pdf}{Github}~\cite{GenDetect_Discovery}, with full details including transaction hashes, similarity scores, and matched references. This enables readers to independently inspect and validate the findings, also promotes transparency. We then analyze these results to understand their structural characteristics and to illustrate the effectiveness of our semantic similarity framework in capturing meaningful but unreported exploit behaviors.

\PP{Setup}
To evaluate our system in a fully realistic setting, we conduct similarity-based detection on real Ethereum transaction data. In theory, the Ethereum mainnet has recorded over 1.5 billion transactions from January 2021 to December 2024. However, to make the evaluation computationally feasible, we restrict our analysis to a targeted subset of transactions involving decentralized exchanges (DEXs), which are more likely to contain financially meaningful operations. This filtered dataset consists of approximately 3 million transactions, collected directly from the Dune Analytics platform~\cite{dune}. The full detection pipeline was executed over this dataset in an end-to-end fashion, with a total processing time of approximately \textbf{77 hours}.

\PP{Results}
Our full detection results include 682 public disclosed incidents, 56 original disclosure, 1862 MEV transactions (excluded from FP by definition) and 79 false positives, yielding an overall 3\% FPR, consistent with cross-validation results. These non-original cases are listed in \href{https://github.com/NobodyIsAnonymous/GenDetect_ICSE2026/blob/main/original_results.txt}{Github Repo}~\cite{GenDetect_original_results}.
To isolate our original disclosures, we applied a rigorous filtering pipeline. First, we excluded arbitrage and MEV-related transactions by analyzing sender history and existing MEV labels from Ethereum datasets\cite{etherscan}. Next, we filtered out known attack incidents using exploiter tags from both Ethereum and Phalcon~\cite{phalcon}. The remaining candidates underwent manual validation based on strong indicators of exploitation: the use of Tornado Cash or cross-chain bridges for rapid fund exfiltration, temporary contract/account creation, and manual confirmation of vulnerabilities in the victim contracts. This conservative process yielded a final set of \textbf{56} previously undocumented attacks~\cite{GenDetect_Discovery}, with over \textbf{\$1.5 million USD} in total financial damage. Unlike tools such as POMABuster, which report any suspicious transactions, our result prioritizes high-confidence real-world attacks with concrete behavioral evidence.


\emph{Case Study and Reference.}
Each representative attack case corresponds to a known historical exploit in DeFiHackLab~\cite{defihacklab}. For brevity, all full-length transaction hashes are listed in the \href{https://github.com/NobodyIsAnonymous/GenDetect_ICSE2026/blob/main/discovered_attacks.pdf}{Github Repo}~\cite{GenDetect_Discovery}, with inline citations provided for verifiability.
Among these, one notable category involves 10 previously unreported attacks that exhibit high structural similarity to the \textit{pSeudoEth} incident. These attacks exploit a mispriced internal accounting mechanism by invoking \texttt{skim()} to redirect funds to the vault itself, resulting in unintended profit. Some individual gains were relatively modest, around \$100, while others exceeded \$3k, reflecting a wide spectrum of exploit severity. Despite this variation, the attacks exhibit consistent behavioral patterns, highlighting the systematic nature of the exploitation. These lower-value cases may have slipped under the radar of existing tools, emphasizing the strength of our approach: by analyzing core transactional logic rather than surface-level attributes, our system remains robust against obfuscation, randomness, and value-based evasion.

\subsection{RQ3: Detection Efficiency}\label{sec:exp_RQ2_novel}

\begin{figure}[ht]
    \centering
    \includegraphics[width=0.90\linewidth]{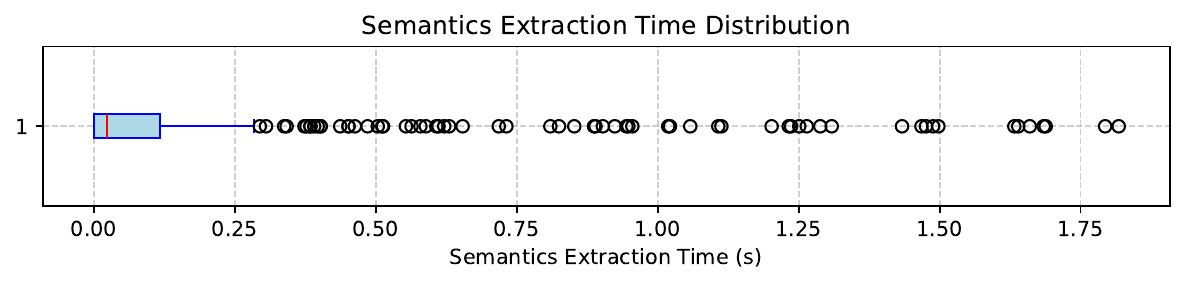}
    \caption{Runtime Distribution of Semantic Extraction. \textit{The box plot illustrates the latency distribution across all 534 attack traces, including median, quartiles, and outliers, reflecting per-transaction variance.}}
    \label{fig:encode_time_boxplot}
\end{figure}

\begin{figure}[ht]
    \centering
    \includegraphics[width=0.90\linewidth]{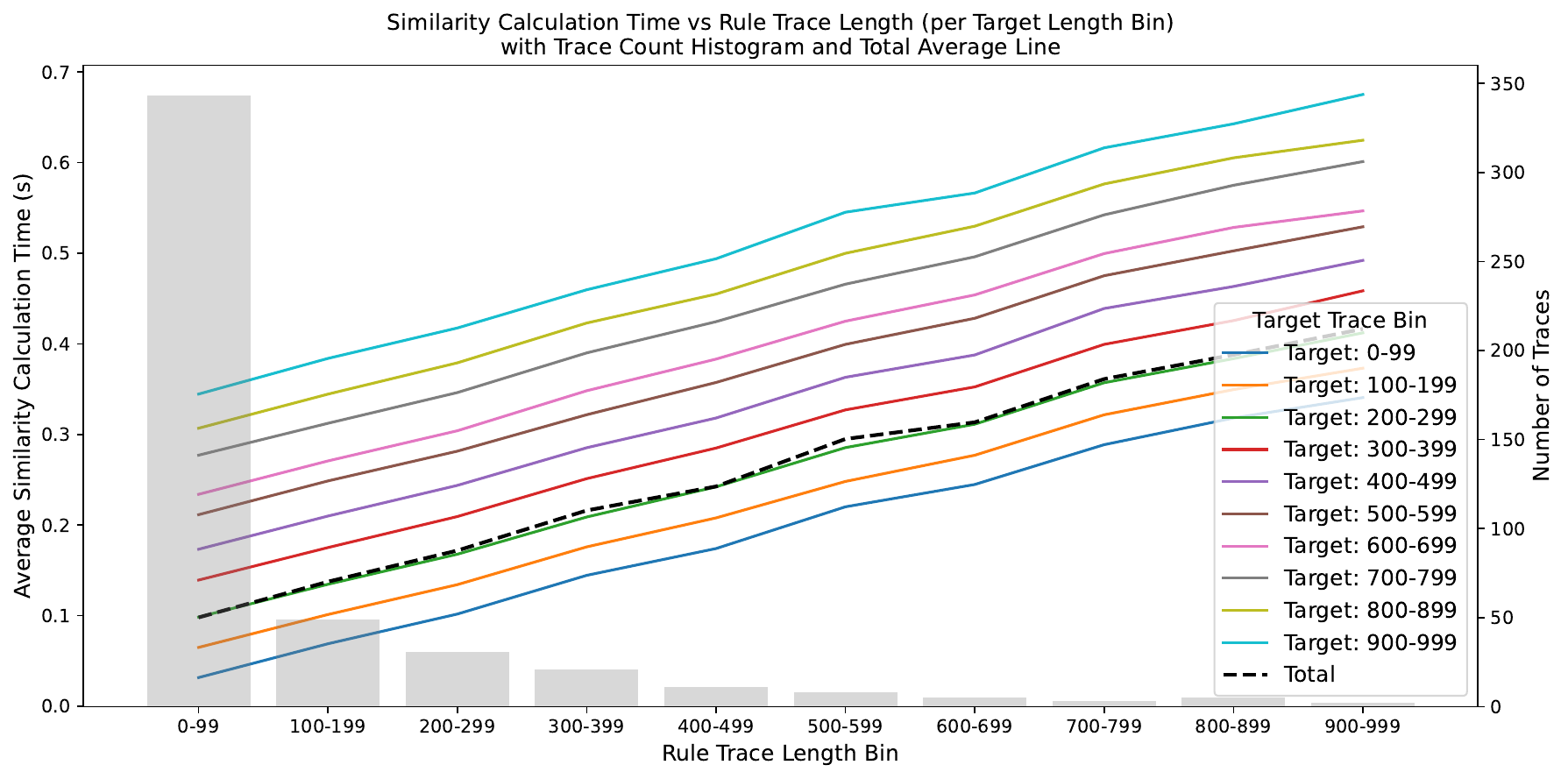}
    \caption{\textbf{Runtime of Logic Extraction and Matching under Varying Rule and Target Trace Lengths.} \textit{Bar Chart: the distribution of rule trace lengths in the dataset; Colored Lines: the average Logic Extraction and Matching time under a specific target trace length; Dashed Line: the global average latency across all target lengths for each rule length.}}
    \label{fig:detection_time}
\end{figure}

In this section, we evaluate the runtime efficiency of our system, with a focus on the two components most critical to real-time performance: the Semantic Cheatsheet Module and the Logic Extraction-Matching Module. We measure the per-transaction processing time of each module under varying trace lengths, demonstrating the practicality of our system in latency-sensitive settings.
The other components in automatic semantic classification and validation pipeline are excluded from detailed timing analysis, because it is executed entirely offline and does not affect online performance. For completeness, we briefly summarize its expected cost profiles.

\PP{Setup}
To evaluate the runtime performance of our system, we measure the processing latency on a dataset of 534 attack transactions collected from DeFiHackLab~\cite{defihacklab}. These transactions represent real-world DeFi exploits and typically exhibit significantly more complex trace structures than regular user transactions. As such, the response times reported in this experiment reflect our system's behavior under \textbf{worst-case conditions}. The primary goal of this evaluation is to assess the feasibility of real-time detection in high-complexity scenarios.

\PP{Results}
Our experiments reveal clear patterns in how runtime scales with trace complexity. For the Semantic Cheatsheet Module (\autoref{fig:encode_time_boxplot}), the average per-transaction latency remains under \textbf{0.02s}, with only a few outliers exceeding \textbf{1.75s}. These rare cases typically involve highly irregular trace structures. For the Logic Module (\autoref{fig:detection_time}), both the length of the rule trace and the target trace jointly influence the runtime. In the worst-case configuration--where both the rule and the target reach the maximum trace length of 800--the observed latency remains bounded around \textbf{0.7s} and the global average runtime remains close to \textbf{0.1s}. These results align well with the theoretical complexity of the logic matching process, which operates at \textbf{$O(M+N)$}, where $M$ and $N$ denote the lengths of the rule and target traces, respectively.


\emph{Real-Time Scenario Feasibility.}
The true performance lower bound only manifests under extremely long traces--cases that are exceedingly rare in practice (\textless0.1\%) and insufficient for statistical characterization. To assess real-time viability, we therefore conduct an analysis using both average throughput and worst-case performance.
\circled{1} For throughput, our system is capable of processing \textbf{over 2,000 transactions per second (TPS)} on 96 cores, based on Ethereum's~\cite{etherscan} transaction length distribution. This throughput comfortably satisfies the real-time detection demands of most blockchain platforms, including Solana~\cite{solana} under average load (1,133 TPS). While Solana's theoretical throughput limit is higher (65,000 TPS), our system is parallelizable by design and can be scaled horizontally to meet such high demands.
\circled{2} Under worst-case performance, for instance, when Ethereum~\cite{etherscan} (\textbf{12s}) and BNB~\cite{Bnb} (\textbf{3s}) encounter one long-trace transaction, whose detection can be fully completed within \textbf{0.7s}. Since only the longest transactions contributes to the \textbf{critical path}, the remaining transactions can be processed in parallel on other cores, allowing total evaluation to finish well within the block interval. Hence, our system satisfies real-time requirements on these platforms. However, on Solana~\cite{solana} (\textbf{0.4s}), GenDetect's worst-case cost slightly exceeds the block interval. We further discuss this limitation and potential solutions in ~\autoref{sec:discussion}.

\section{Limitation and Discussion}\label{sec:discussion}

Our approach presents two primary limitations: real-time performance on high-speed chains and evasion robustness.

\PP{Performance} While the method demonstrates low-latency detection per trace, meeting extreme throughput demands (e.g., Solana's 65K TPS) would require impractical horizontal scaling, approximately 30$\times$ more cores. A more feasible path is to pre-filter trivial transactions (e.g., those with $\leq 2$ calls), which make up ~99.7\% of total traffic. With only ~86 long transactions per 0.4s block interval, and 0.1s average analysis time, we estimate that 22 cores suffice to meet practical real-time needs. Even for worst-case traces (0.7s), our Asymmetrical Normalized Set Difference (ANSD) algorithm supports parallel computation, where doubling the cores reduces worst-case latency to ~0.35s, safely within a block interval.

\PP{Potential Evasions}
Although GenDetect exhibits robustness against post-hoc mutation and noise, several potential evasion vectors remain. These can be categorized into three types below, each reflecting a different layer of the detection pipeline.

\emph{First-instance Evasion:} Our contract-labelling and ANSD design mitigates most post-hoc evasions by filtering non-critical semantic mutations. However, if attackers anticipate our theory and inject obfuscation at the first occurrence of an attack pattern, the initial representation itself becomes noisy, degrading downstream similarity learning. Formalizing and defending against such first-instance evasions remain open challenges for current pattern-based systems.

\emph{Label-compromise Evasion:} Because GenDetect relies on a manually curated address-label dataset, an attacker could theoretically evade detection by compromising the labelling process and having malicious addresses marked as benign. This would bypass address-based filtration. In practice, such side-channel manipulation is difficult since the label set is manually verified and periodically maintained, but it highlights the importance of data-source integrity.

\emph{Private-relay Evasion:} Private Relay Network is originally introduced by Flashbots~\cite{flashbot} to mitigate MEV by bypassing the public mempool. Some permissioned relays expose authenticated APIs that authorized institutions can monitor, allowing GenDetect to extend to these environments by incorporating relay or bundle data alongside public-mempool captures.
However, fully closed private relays (often referred to as "dark-pools") pose a greater challenge. Transactions submitted exclusively through such relays are invisible to public-mempool monitors before inclusion, placing them outside the observable scope of GenDetect and other mempool-based detectors. Detecting attacks launched purely via closed private relays thus remains an open problem and represents an inherent limitation of observation-based detection. Future directions include deeper collaboration with relay operators, on-chain trace correlation, and new visibility instrumentation.

\section{Related Work}
\label{sec:related_work}

A variety of tools have been developed to detect malicious transactions in DeFi, primarily following pattern-based or contract-level detection paradigms. Pattern-based systems such as DeFiRanger~\cite{wu2023defiranger} and POMABuster~\cite{POMABuster_code} identify price-driven attacks by matching trade and oracle manipulation patterns, while TxSpector~\cite{txSpector_code} targets logic bugs like reentrancy and unchecked external calls. Other domain-specific tools include LeiShen~\cite{xia2023LeiShen} for flashloan-related attacks and TLMG4Eth~\cite{sun2025TLMG4Eth} for address-level fraud detection.
Contract-level tools like Forta~\cite{Forta} analyze deployed bytecode to flag malicious behavior before execution, but face challenges due to compilation variance and limited semantic interpretability.
Some works address post-exploit defense: APE~\cite{qin2023APE} and Sting~\cite{zhang2023Sting} simulate or mimic known attacks, while FlashGuard~\cite{alhaidari2025nonPrice} mitigates non-price vulnerabilities after the fact. In contrast, our work focuses on generalizing detection patterns from traces to enable real-time, semantic-level detection of novel and polymorphic DeFi attacks.

\section{Conclusion}
\label{sec:conclusion}
This paper presents GenDetect, a detection generation framework designed to address the persistent challenge of recurring DeFi attacks through semantic generalization and structural similarity analysis. By abstracting execution traces into financial semantics and comparing them compositionally, our system captures the underlying intent of transactions rather than relying on brittle surface-level features. Its fully parallelizable per-transaction design--along with a set-based similarity metric (ANSD) that is inherently parallelizable--enables high-throughput performance, making it suitable for real-time deployment even under modern blockchain traffic. Our evaluation demonstrates its effectiveness across both benchmark and zero-shot settings, while uncovering previously unreported attacks in the wild. As DeFi continues to evolve, we believe GenDetect offers a robust foundation for scalable, interpretable, and timely threat detection, and we envision future work exploring enhanced semantic representations and accelerated similarity computation to further strengthen real-time defenses.

\section{Acknowledgement}
Kangjie Lu, Bowen Cai and Weiheng Bai were supported in part by NSF awards CNS-2045478, CNS-2106771, and CNS-2247434. Any opinions, findings, conclusions or recommendations expressed in this material are those of the authors and do not necessarily reflect the views of NSF.


\bibliographystyle{sty/ACM-Reference-Format.bst}
\setlength{\bibsep}{3pt}
\bibliography{mybib}

@string{ICDCS     = { International Conference on Distributed Computing Systems (ICDCS)}}

@string{SP        = { IEEE Symposium on Security and Privacy (Oakland)}}

@string{SEC       = { USENIX Security Symposium (Security)}}

@string{CRYPTO    = { International Cryptology Conference (CRYPTO)}}

@misc{dex_trades,
  author       = {Dune Analytics},
  title        = {Dune Dex Trades Data},
  url          = {https://www.dune.com/data/dex.trades},
  abstractNote = {Dune is the all-in-one crypto data platform — query with SQL, stream data via APIs \& DataShare, and publish interactive dashboards across 100+ blockchains.},
  language     = {en},
  year         = {2025}
}

@misc{dex_aggregator_trades,
  author       = {Dune Analytics},
  title        = {Dune Dex Aggregator Trades Data},
  url          = {https://www.dune.com/data/dex_aggregator.trades},
  abstractNote = {Dune is the all-in-one crypto data platform — query with SQL, stream data via APIs \& DataShare, and publish interactive dashboards across 100+ blockchains.},
  language     = {en},
  year         = {2025}
}

@misc{nft_trades,
  author       = {Dune Analytics},
  title        = {Dune NFT Trades Data},
  url          = {https://www.dune.com/data/nft.trades},
  abstractNote = {Dune is the all-in-one crypto data platform — query with SQL, stream data via APIs \& DataShare, and publish interactive dashboards across 100+ blockchains.},
  language     = {en},
  year         = {2025}
}

@misc{staking_ethereum_flows,
  author       = {Dune Analytics},
  title        = {Dune Staking Ethereum Flows Data},
  url          = {https://www.dune.com/data/staking_ethereum.flows},
  abstractNote = {Dune is the all-in-one crypto data platform — query with SQL, stream data via APIs \& DataShare, and publish interactive dashboards across 100+ blockchains.},
  language     = {en},
  year         = {2025}
}

@misc{flashbot,
  author       = {Flashbots},
  title        = {Flashbots MEV-Boost},
  url          = {https://boost.flashbots.net},
  abstractNote = {MEV-Boost is an implementation of proposer-builder separation (PBS) built by Flashbots for proof of stake Ethereum. Validators running MEV-Boost maximize their staking reward by selling blockspace to an open market of builders.},
  journal      = {MEV-Boost in a Nutshell},
  language     = {en},
  year         = {2025}
}

@article{wu2023defiranger,
  title     = {Defiranger: detecting DeFI price manipulation attacks},
  author    = {Wu, Siwei and Yu, Zhou and Wang, Dabao and Zhou, Yajin and Wu, Lei and Wang, Haoyu and Yuan, Xingliang},
  journal   = {IEEE Transactions on Dependable and Secure Computing},
  volume    = {21},
  number    = {4},
  pages     = {4147--4161},
  year      = {2023},
  publisher = {IEEE}
}

@inproceedings{xi2024pomabuster,
  title        = {POMABuster: Detecting Price Oracle Manipulation Attacks in Decentralized Finance},
  author       = {Xi, Rui and Wang, Zehua and Pattabiraman, Karthik},
  booktitle    = {2024 IEEE Symposium on Security and Privacy (SP)},
  pages        = {3923--3942},
  year         = {2024},
  organization = {IEEE},
  address      = {San Francisco, CA, USA},
  publisher = {IEEE},
}

@article{sun2025TLMG4Eth,
  title     = {Ethereum fraud detection via joint transaction language model and graph representation learning},
  author    = {Sun, Jianguo and Jia, Yifan and Wang, Yanbin and Tian, Ye and Zhang, Sheng},
  journal   = {Information Fusion},
  volume    = {120},
  pages     = {103074},
  year      = {2025},
  publisher = {Elsevier}
}

@misc{Bnb,
  title        = {BNB Smart Chain (BNB) Blockchain Explorer},
  url          = {https://bscscan.com/},
  abstractnote = {BscScan allows you to explore and search the Binance blockchain for transactions, addresses, tokens, prices and other activities taking place on Binance (BNB)},
  journal      = {BNB Smart Chain Explorer},
  author       = {BscScan.com},
  language     = {en},
  year = {2025},
}

@misc{IndustryApproach,
  title        = {MetaSuites (formerly MetaDock): The Builder's Swiss Army Knife by BlockSec},
  url          = {https://blocksec.com/metasuites},
  abstractnote = {Generate fund flow, display address labels, download data with one click, simulate transactions, view storage, and proxy upgrades on over ten blockchain explorers.},
  journal      = {BlockSec},
  author       = {BlockSec},
  language     = {en},
  year = {2025},
}

@inproceedings{xia2023LeiShen,
  title        = {Detecting Flash Loan Based Attacks in Ethereum},
  author       = {Xia, Qing and Huang, Zhirong and Dou, Wensheng and Zhang, Yafeng and Zhang, Fengjun and Liang, Geng and Zuo, Chun},
  booktitle    = {2023 IEEE 43rd International Conference on Distributed Computing Systems (ICDCS)},
  pages        = {154--165},
  year         = {2023},
  organization = {IEEE},
  address      = {Online},
  publisher = {IEEE},
}

@inproceedings{zhang2020txspector,
  title     = {$\{$TXSPECTOR$\}$: Uncovering attacks in ethereum from transactions},
  author    = {Zhang, Mengya and Zhang, Xiaokuan and Zhang, Yinqian and Lin, Zhiqiang},
  booktitle = {29th USENIX Security Symposium (USENIX Security 20)},
  pages     = {2775--2792},
  year      = {2020}
}

@inproceedings{qin2023APE,
  title     = {The blockchain imitation game},
  author    = {Qin, Kaihua and Chaliasos, Stefanos and Zhou, Liyi and Livshits, Benjamin and Song, Dawn and Gervais, Arthur},
  booktitle = {32nd USENIX Security Symposium (USENIX Security 23)},
  pages     = {3961--3978},
  year      = {2023}
}

@article{mabruri2024dynamic,
  title   = {Dynamic Taxonomy a Bridge from DeFi to TradFi},
  author  = {Mabruri, Iqbal and Cerqueda, Carles and Jack, Christopher and Lui, Alexis and Wu, Wenbin and DiPerna, Vincezo and Bear, Keith and Zhang, Bryan and others},
  journal = {Available at SSRN},
  year    = {2024}
}

@article{gogol2024sok,
  title   = {SoK: Decentralized Finance (DeFi)--Fundamentals, Taxonomy and Risks},
  author  = {Gogol, Krzysztof and Killer, Christian and Schlosser, Malte and Bocek, Thomas and Stiller, Burkhard and Tessone, Claudio},
  journal = {arXiv preprint arXiv:2404.11281},
  year    = {2024}
}

@inproceedings{xu2022short,
  title        = {A short survey on business models of decentralized finance (DeFi) protocols},
  author       = {Xu, Teng Andrea and Xu, Jiahua},
  booktitle    = {International Conference on Financial Cryptography and Data Security},
  pages        = {197--206},
  year         = {2022},
  organization = {Springer}
}

@inproceedings{moncada2021next,
  title        = {Next generation blockchain-based financial services},
  author       = {Moncada, Roberto and Ferro, Enrico and Favenza, Alfredo and Freni, Pierluigi},
  booktitle    = {Euro-Par 2020: Parallel processing workshops: Euro-Par 2020 international workshops, Warsaw, Poland, August 24--25, 2020, Revised selected papers 26},
  pages        = {30--41},
  year         = {2021},
  organization = {Springer}
}

@article{aquilina2024decentralized,
  title     = {Decentralized finance (DeFi): a functional approach},
  author    = {Aquilina, Matteo and Frost, Jon and Schrimpf, Andreas},
  journal   = {Journal of Financial Regulation},
  volume    = {10},
  number    = {1},
  pages     = {1--27},
  year      = {2024},
  publisher = {Oxford University Press}
}

@misc{phalcon,
  author   = {BlockSec},
  title    = {Security {Incidents} {\textbar} {Phalcon} {Explorer}},
  url      = {https://app.blocksec.com},
  language = {en},
  urldate  = {2025-04-19}
}

@misc{defihacklab,
  title    = {{SunWeb3Sec}/{DeFiHackLabs}},
  url      = {https://github.com/SunWeb3Sec/DeFiHackLabs},
  urldate  = {2025-04-19},
  author   = {SunWeb3Sec},
  month    = apr,
  year     = {2025},
  note     = {original-date: 2022-06-10T09:57:11Z}
}

@misc{Cymetrics,
  author       = {AliceHsu},
  url          = {https://tech-blog.cymetrics.io/en/posts/alice/2024_defi_hack/},
  abstractnote = {In the first half of 2024, significant progress was made in the cryptocurrency space. The U.S. Securities and Exchange Commission (SEC) approved a spot Bitcoin ETF in January, followed by the approval of a spot Ethereum ETF in July.Meanwhile, on April 23, the Hong Kong Securities and Futures Commission (SFC) approved three asset management companies-Bosera International, China Asset Management, and Harvest Global Investments-to issue spot Bitcoin and Ethereum ETFs. These products were officially listed on exchanges on April 30.The DeFi sector subsequently experienced diversified and vibrant development. Driven by market anticipation of regulatory policies, Bitcoin surpassed the $100,000 mark. According to statistics from DeFiLlama, on January 1, 2024, the total value locked (TVL) in DeFi stood at $54.162 billion. As of now (December 2024), the TVL has risen significantly, exceeding $100 billion.These changes symbolize the integration of crypto assets into traditional financial markets, bringing more diversified services, investment opportunities, and potential returns.},
  journal      = {Cymetrics Tech Blog},
  language     = {en}
}

@misc{Forta,
  title        = {Forta},
  url          = {https://www.forta.org/},
  abstractnote = {Powered by the most advanced AI detection model, Forta Firewall integrates with protocols and rollups to prevent over 99% of hacks.},
  author       = {Forta},
  language     = {en}
}

@article{nadler2023tornado,
  title   = {Tornado cash and blockchain privacy: a primer for economists and policymakers},
  author  = {Nadler, Matthias and Sch{\"a}r, Fabian},
  journal = {Federal Reserve Bank of St. Louis Review},
  year    = {2023}
}

@misc{recur_1,
  title        = {0x5e694707337cca979d | Phalcon Explorer},
  url          = {https://app.blocksec.com},
  abstractnote = {Security suite for Protocols, Developers, LPs, & Traders - safeguarding your blockchain journey},
  author       = {BlockSec},
  language     = {en}
}

@misc{recur_2,
  title        = {0x75e3aeb00df69882a1 | Phalcon Explorer},
  url          = {https://app.blocksec.com},
  abstractnote = {Security suite for Protocols, Developers, LPs, & Traders - safeguarding your blockchain journey},
  author       = {BlockSec},
  language     = {en}
}

@misc{recur_idol_nft,
  title        = {0xa8289dbe3e49e9c5de | Phalcon Explorer},
  url          = {https://app.blocksec.com},
  abstractnote = {Security suite for Protocols, Developers, LPs, & Traders - safeguarding your blockchain journey},
  author       = {BlockSec},
  language     = {en}
}

@misc{recur_4,
  title        = {0xbeefd8faba2aa82704 | Phalcon Explorer},
  url          = {https://app.blocksec.com},
  abstractnote = {Security suite for Protocols, Developers, LPs, & Traders - safeguarding your blockchain journey},
  author       = {BlockSec},
  language     = {en}
}

@misc{4bytes_contract,
  title  = {ethereum-lists/contracts: List of contracts from known projects (work in progress)},
  url    = {https://github.com/ethereum-lists/contracts},
  author = {4bytes},
  year = {2025}
}

@misc{GenDetect_original_results,
  title        = {GenDetect Repository: Complementary Original Results Data},
  url          = {https://github.com/NobodyIsAnonymous/GenDetect_ICSE2026/blob/main/original_results.txt},
  journal      = {GitHub},
  author       = {{GenDetect Authors}},
  language     = {en},
  year         = {2025}
}

@misc{GenDetect_Discovery,
  title  = {GenDetect Repository: Newly Discovered Attacks Data},
  url    = {https://github.com/NobodyIsAnonymous/GenDetect_ICSE2026/blob/main/discovered_attacks.pdf},
  author = {{GenDetect Authors}},
  year   = {2025}
}

@misc{GenDetect_Zero-shot_DeFiHackLab,
  title        = {GenDetect Repository: Cross-domain Zero-shot Evaluation Data (DeFiHackLab)},
  url          = {https://github.com/NobodyIsAnonymous/GenDetect_ICSE2026/blob/main/DeFiHackLab_zero_test.csv},
  journal      = {GitHub},
  author       = {{GenDetect Authors}},
  language     = {en},
  year         = {2025}
}

@misc{GenDetect_Zero-shot,
  title        = {GenDetect Repository: Complementary Zero-shot Evaluation Data},
  url          = {https://github.com/NobodyIsAnonymous/GenDetect_ICSE2026/blob/main/zero-shot_evaluation.pdf},
  journal      = {GitHub},
  author       = {{GenDetect Authors}},
  language     = {en},
  year         = {2025}
}

@article{attack_survey,
  title     = {Exploring the attack surface of blockchain: A comprehensive survey},
  author    = {Saad, Muhammad and Spaulding, Jeffrey and Njilla, Laurent and Kamhoua, Charles and Shetty, Sachin and Nyang, DaeHun and Mohaisen, David},
  journal   = {IEEE Communications Surveys \& Tutorials},
  volume    = {22},
  number    = {3},
  pages     = {1977--2008},
  year      = {2020},
  publisher = {IEEE}
}

@inproceedings{xia2023flashLoan,
  title        = {Detecting Flash Loan Based Attacks in Ethereum},
  author       = {Xia, Qing and Huang, Zhirong and Dou, Wensheng and Zhang, Yafeng and Zhang, Fengjun and Liang, Geng and Zuo, Chun},
  booktitle    = {2023 IEEE 43rd International Conference on Distributed Computing Systems (ICDCS)},
  pages        = {154--165},
  year         = {2023},
  organization = {IEEE}
}

@inproceedings{mo2023priceFeed,
  title     = {Toward automated detecting unanticipated price feed in smart contract},
  author    = {Mo, Yifan and Chen, Jiachi and Wang, Yanlin and Zheng, Zibin},
  booktitle = {Proceedings of the 32nd ACM SIGSOFT International Symposium on Software Testing and Analysis},
  pages     = {1257--1268},
  year      = {2023}
}

@inproceedings{wang2021promutator,
  title        = {ProMutator: Detecting vulnerable price oracles in DeFi by mutated transactions},
  author       = {Wang, Shih-Hung and Wu, Chia-Chien and Liang, Yu-Chuan and Hsieh, Li-Hsun and Hsiao, Hsu-Chun},
  booktitle    = {2021 IEEE European Symposium on Security and Privacy Workshops (EuroS\&PW)},
  pages        = {380--385},
  year         = {2021},
  organization = {IEEE}
}

@article{alhaidari2025nonPrice,
  title   = {Protecting DeFi Platforms against Non-Price Flash Loan Attacks},
  author  = {Alhaidari, Abdulrahman and Palanisamy, Balaji and Krishnamurthy, Prashant},
  journal = {arXiv preprint arXiv:2503.01944},
  year    = {2025}
}

@misc{etherscan,
  title        = {Ethereum (ETH) Blockchain Explorer},
  url          = {https://etherscan.io/},
  abstractnote = {Etherscan allows you to explore and search the Ethereum blockchain for transactions, addresses, tokens, prices and other activities taking place on Ethereum (ETH)},
  journal      = {Ethereum (ETH) Blockchain Explorer},
  author       = {etherscan.io},
  language     = {en}
}

@misc{solana,
  title        = {Explorer | Solana},
  url          = {https://explorer.solana.com/},
  abstractnote = {Inspect transactions, accounts, blocks, and more on the Solana blockchain},
  author       = {solana},
  language     = {en}
}

@article{katsiampa2019volatility,
  title     = {Volatility co-movement between Bitcoin and Ether},
  author    = {Katsiampa, Paraskevi},
  journal   = {Finance Research Letters},
  volume    = {30},
  pages     = {221--227},
  year      = {2019},
  publisher = {Elsevier}
}

@inproceedings{ji2021reentrancy,
  title        = {Security Analysis of Blockchain Smart Contract: Taking Reentrancy Vulnerability as an Example},
  author       = {Ji, Mingtao and Liang, GuangJun and Li, Meng and Zhang, Haoyan and He, Jiacheng},
  booktitle    = {Advances in Artificial Intelligence and Security: 7th International Conference, ICAIS 2021, Dublin, Ireland, July 19-23, 2021, Proceedings, Part III 7},
  pages        = {492--501},
  year         = {2021},
  organization = {Springer}
}

@article{wu2023privilege,
  title     = {TaintGuard: Preventing implicit privilege leakage in smart contract based on taint tracking at abstract syntax tree level},
  author    = {Wu, Xiangyu and Du, Xuehui and Yang, Qiantao and Liu, Aodi and Wang, Na and Wang, Wenjuan},
  journal   = {Journal of Systems Architecture},
  volume    = {141},
  pages     = {102925},
  year      = {2023},
  publisher = {Elsevier}
}

@inproceedings{zhang2024accounting,
  title     = {Towards finding accounting errors in smart contracts},
  author    = {Zhang, Brian},
  booktitle = {Proceedings of the IEEE/ACM 46th International Conference on Software Engineering},
  pages     = {1--13},
  year      = {2024}
}

@misc{codebert,
  author       = {Feng, Zhangyin and Guo, Daya and Tang, Duyu and Duan, Nan and Feng, Xiaodong and Gong, Ming and Shou, Linjun and Qin, Bing and Liu, Ting and Jiang, Daxin and Zhou, Ming},
  title        = {{CodeBERT: A Pre-Trained Model for Programming and Natural Languages}},
  year         = {2020},
  howpublished = {\url{https://github.com/microsoft/CodeBERT}},
  note         = {arXiv:2002.08155}
}

@inproceedings{zhang2023Sting,
  title     = {Your exploit is mine: Instantly synthesizing counterattack smart contract},
  author    = {Zhang, Zhuo and Lin, Zhiqiang and Morales, Marcelo and Zhang, Xiangyu and Zhang, Kaiyuan},
  booktitle = {32nd USENIX Security Symposium (USENIX Security 23)},
  pages     = {1757--1774},
  year      = {2023}
}

@misc{dune,
  author       = {Dune Analytics},
  title        = {Dune: On-chain Crypto Data and Analytics},
  year         = {n.d.},
  howpublished = {\url{https://dune.com}},
  note         = {Accessed April 19, 2025}
}

@misc{Forta_code,
  type      = {Jupyter Notebook},
  title     = {forta-network/starter-kits},
  rights    = {MIT},
  url       = {https://github.com/forta-network/starter-kits},
  publisher = {Forta Network},
  author    = {Forta},
  year      = {2025},
  month     = apr
}

@misc{txSpector_code,
  title  = {OSUSecLab/TxSpector},
  url    = {https://github.com/OSUSecLab/TxSpector/tree/master},
  author = {OSUSecLab}
}

@misc{POMABuster_code,
  title  = {DependableSystemsLab/POMABuster: POMABuster is an automated engine to detect Price Oracle Manipualtion Attack (POMA) to blockchain oracles.},
  url    = {https://github.com/DependableSystemsLab/POMABuster/tree/main},
  author = {Siriussee}
}

@article{hagele2024DEX,
  title     = {Centralized exchanges vs. decentralized exchanges in cryptocurrency markets: A systematic literature review},
  author    = {H{\"a}gele, Sascha},
  journal   = {Electronic Markets},
  volume    = {34},
  number    = {1},
  pages     = {33},
  year      = {2024},
  publisher = {Springer}
}

@article{schar2021Derivatives,
  title     = {Decentralized finance: on blockchain and smart contract-based financial markets},
  author    = {Sch{\"a}r, Fabian},
  journal   = {Review of the Federal Reserve Bank of St Louis},
  volume    = {103},
  number    = {2},
  pages     = {153--174},
  year      = {2021},
  publisher = {Federal Reserve Bank of St. Louis}
}

@misc{MinerToken,
  title        = {0x75e3aeb00df69882a1 | Phalcon Explorer},
  url          = {https://app.blocksec.com},
  abstractnote = {Security suite for Protocols, Developers, LPs, & Traders - safeguarding your blockchain journey},
  author       = {Phalcon},
  language     = {en}
}


\end{document}